\documentclass[authoryear-refs]{wiley-article}

\usepackage{siunitx}
\usepackage[natbibapa]{apacite}
\usepackage{setspace}
\usepackage{lipsum}
\usepackage{booktabs}

\papertype{Original Article}
\paperfield{Journal Section}

\title{From Retrieval to Synthesis: Repair Literacy and the Domestication of Generative AI}

\abbrevs{AI, Artificial Intelligence; GenAI, Generative Artificial Intelligence; ISP, Information Search Process; IB, Information Behavior; UGT, Uses and Gratifications Theory; SDL, Self-Directed Learning.}

\author[1]{Tawfiq Ammari}
\author[1]{Meilun Chen}
\author[1]{S M Mehedi Zaman}
\author[1]{Kiran Garimella}

\affil[1]{Rutgers University School of Communication and Information, 4 Huntington Street, New Brunswick, NJ 08901, USA}

\corraddress{Tawfiq Ammari, Rutgers University School of Communication and Information, 4 Huntington Street, New Brunswick, NJ 08901, USA}
\corremail{tawfiq.ammari@rutgers.edu}
\fundinginfo{}
\runningauthor{Ammari et al.}

\begin{document}

\begin{frontmatter}
\maketitle

\begin{abstract}
How do students develop AI literacy through everyday practice rather than formal instruction? While normative AI literacy frameworks proliferate, empirical understanding of how students actually learn to work with generative AI remains limited. This study analyzes 10,536 ChatGPT messages from 36 undergraduates over one academic year, revealing five use genres (\textit{Workhorse}, \textit{Algorithmic Auditing and Repair}, \textit{Anthropomorphic Sensemaking}, \textit{Metacognitive Processing}, and \textit{Trust Calibration}) that constitute distinct configurations of student--AI learning. We use the term \textit{repair literacy} to describe the competencies students develop through diagnosing, negotiating, and recovering from AI failures, a crucial but underexplored dimension of AI competence that emerges through practice rather than formal instruction. Drawing on domestication theory and information behavior frameworks, we demonstrate that functional AI competence emerges through ongoing relational negotiation. This process reframes information seeking not as simple retrieval, but as a complex sense-making loop where students continuously calibrate trust and verify algorithmic outputs. Students develop sophisticated genre portfolios, strategically matching interaction patterns to learning needs while exercising critical judgment about AI limitations. Our findings offer information scientists empirically grounded insights into how students actually learn to work with generative AI, with implications for algorithmic folk theories, responsible AI integration, and the design of AI-enabled learning environments that support student agency.

\keywords{AI literacy, self-regulated learning, student agency, generative AI, ChatGPT, human-AI interaction, responsible AI}
\end{abstract}
\end{frontmatter}




\section{Introduction}
Generative AI has moved rapidly into students' academic lives, raising an urgent question: how do students develop the competence to work with these systems? AI-literacy frameworks have proliferated \citep{ng2021conceptualizing,long2020ai} and intervention studies report knowledge gains \citep{laupichler2022evaluation} and sharper critical thinking \citep{essel2024chatgpt}, yet we know little about how such competence forms through naturalistic, everyday practice. This study takes up that gap empirically.

Information science has long treated seeking as active and constructive. In Bates' berrypicking the query evolves as each document reshapes the need, met through small selections gathered one at a time \citep{bates1989design}. Kuhlthau's Information Search Process casts this as multi-stage inquiry, cognitive, affective, and behavioral at once, in which the searcher synthesizes what they gather into a point of view \citep{kuhlthau1991inside}. A conversational model now joins the forage, returning finished prose and shifting the user from \emph{search as retrieval} to \emph{search as synthesis} \citep{sebastian2025reframing}. As an \emph{information intermediary} it composes rather than selects \citep{guner2025ai,shah_bender_22}, taking on part of the synthetic labor so the work becomes refining prompts, not chasing documents. Dervin casts information behavior as gap-bridging: at a moment of uncertainty the user builds a bridge across a gap, so sense is constructed through situated observing rather than found in the world \citep{dervin1983information}. An inadequate response is simply a fresh gap, and the repair that follows, rephrasing, supplying context, restarting, does the bridging. In essence, a datafied output means little until the person articulates it against context and lived experience \citep{lupton2018data}, work that cannot be handed to the machine. Metacognition regulates the loop, monitoring understanding and signaling when to probe, accept, or restart \citep{flavell1979metacognition}.

Foraging alongside a generative model has a catch: the seeker cannot see what her companion is doing, and the model can return fluent, confident text that is wrong. \citet{touretzky2023machine} contrast ``black box'' tools with ``glass box'' demonstrations that expose how a system reaches its output, arguing competence means seeing past surface fluency. Lacking that visibility, users improvise \emph{algorithmic folk theories}, informal and often inaccurate models that nonetheless guide interaction and recovery during breakdowns \citep{karizat_et_al_algo_folk_21,ridley2024informing}. Because the companion is opaque and fallible, productive use depends on what we term \emph{repair literacy}: the competencies students develop to diagnose, negotiate, and recover from AI failures.

Repair literacy complements existing AI-literacy frameworks. \citet{ng2021conceptualizing} organize AI competence into cognitive, affective, behavioural, and ethical dimensions \citep{ng2024design}, which recent work extends by threading ethics through every dimension, so that interrogating training-data bias becomes part of cognitive competence \citep{ma2025fostering}, while \citet{jia2025developing} define literacy as understanding AI's characteristics and workings together with the ability to engage responsibly. Repair literacy supplies the in-the-moment competence these models imply but leave underspecified. The point is clearest in Touretzky and colleagues' ``Natural Interaction'' idea, that agents need many kinds of knowledge to interact naturally with people \citep[p.~248]{touretzky2023machine}. The converse also holds. Students must build tacit knowledge of the \emph{AI}. They learn its quirks through the trial-and-error probing which \citet{jarrahi2025interviewing} calls exploratory familiarization. The model itself leans on this repair work to function \citep{putz2024performance}. As answers fall short, new gaps open \citep{dervin1983information}. So bridging strategies must stay equally dynamic.

Repair literacy is rarely taught; it accrues through use. Naturalistic AI use is experimentation in the wild, students varying prompts and accumulating a working theory of the model's behaviour \citep[p.~236]{touretzky2023machine}; what turns this into durable competence is domestication, the folding of ChatGPT into the rhythms of academic life. This matters because AI can prompt students to ``rely on rather than learn from'' it, compromising agency and self-regulation \citep{darvishi2024impact}. Rather than treat ChatGPT as a static tool, we examine how students bring it ``into the distinct cultures'' of academic life \citep[p.~15]{silverstone1992information}, an ongoing negotiation in which users, technologies, and contexts mutually shape one another \citep{sorensen2006domestication}. This is \citeauthor{sorensen2006domestication}'s ``appropriation work,'' in which users fold ChatGPT into their ``moral economy'' \citep[p.~236]{silverstone2006domesticating} and discipline the technology into what \cite{bakardjieva2005internet} calls ``use genres.'' 

Analyzing 10,536 actual interaction messages rather than self-reports, we identify five empirically-derived \textit{use genres} \citep{bakardjieva2005internet} that form the analytic spine of our Findings (§\ref{sec:findings})---the \textit{Workhorse (\S\ref{sec:workhorse})}, \textit{Anthropomorphic Sensemaking (\S\ref{sec:companion})}, \textit{Algorithmic Auditing and Repair (\S\ref{sec:repair})}, \textit{Metacognitive Processing (\S\ref{sec:metacognitive})}, and \textit{Trust Calibration (\S\ref{sec:trust})}---with a single user moving fluidly across them within a semester. By foregrounding situated use over controlled experiments, this study answers calls to examine ``AI in situated action'' \citep{jarrahi2025interviewing} and shows how AI literacy develops as a sociomaterial enactment.

\section{Related Work}

Our review draws together three strands that frame how students learn to work with generative AI: the metacognitive and sense-making processes through which people bridge knowledge gaps, the information-behavior scholarship now being reconfigured by conversational systems, and the AI-literacy frameworks that specify what such competence entails.

\subsection{Information Seeking, Sense-Making, and Metacognition}

Long before generative AI, information scholarship treated seeking as active and constructive rather than simple retrieval, with metacognition as its regulatory layer. Defined as a learner's knowledge about and regulation of their own cognition \citep{flavell1979metacognition}, metacognition distinguishes cognitive strategies, which make progress on a task, from metacognitive strategies, which monitor it. This monitoring lens is now central to accounts of seekers who set goals, track comprehension, and adjust strategy mid-task \citep{kuhlthau1991inside,garrison1997self}.

Such regulation is exercised as people make sense of uncertainty. Dervin's \citeyearpar{dervin1983information} sense-making frames seeking as gap-bridging: at a moment of uncertainty an individual confronts a cognitive-affective gap and builds a bridge across it, where information is not external to the person but constructed through situated observing. The metaphor maps directly onto generative AI, where an inadequate response is a gap and repair moves such as rephrasing, escalating, or restarting are bridge-building. Kuhlthau's Information Search Process \citep{kuhlthau1991inside} formalizes this view as a multi-stage inquiry integrating cognitive, affective, and behavioral dimensions, one that already encompasses synthesis rather than one-off retrieval, where uncertainty eases at the ``formulation'' turning point and seeking is recursive rather than linear. Bates' \citeyearpar{bates1989design} berrypicking model adds that each encountered document reshapes the query itself, so seeking proceeds through a series of selections drawing on a wide repertoire of tactics.

Generative AI redistributes rather than displaces this labor: synthesis is partly delegated to an algorithmic intermediary, and the berrypicking logic inverts, since responses are generated in reply to the query and evolve through prompt refinement rather than document discovery. Because the system behaves less like a tool than an interlocutor, students probe its boundaries through what \cite{jarrahi2025interviewing} call exploratory familiarization, treating the AI assistant as a ``role-maker'' rather than ``role-taker'' \citep{schulz2025generative} in a relationship of human-AI complementarity \citep{holstein2019designing}. When responses break down they engage in resilient repair \citep{Ashktorab_2019}, interpretive work that reflects Lupton's data sense-making. For Lupton a datafied output means little on its own and comes to matter only once a person articulates it, joining it to context and lived experience so that it makes sense. That labor cannot be handed to the machine. Students stay agential, supplying the context the reply cannot furnish on its own and intervening in its claims rather than taking them at face value. In this way they \emph{become with} the system, remaking the AI and being remade by it, so that human-data assemblages are continually recomposed and knowing and being stay entangled \citep{lupton2018data}. An artifact becomes ``working technology'' only once users settle an interpretation \citep{pinch1984social,latour1993pasteurization}. These constructs anticipate two genres: \textit{Metacognitive Processing} (\S\ref{sec:metacognitive}), where ChatGPT is a cognitive mirror for self-explanation rather than retrieval, and \textit{Algorithmic Auditing and Repair} (\S\ref{sec:repair}), where gap-bridging after failure becomes productive sense-making.

\subsection{Trust and Cognitive Authority in Generative AI}

Generative AI has prompted a re-evaluation of information seeking. \cite{sebastian2025reframing} argue that it shifts the user's role from \textit{search as retrieval} to \textit{search as synthesis}, as selecting among sources gives way to prompting and refining generated content, which \cite{greussing2025perception} characterize as positioning generative AI as an information intermediary that reshapes the search process. This changes how users engage. \cite{kim2025consumer} find that users value the conversational quality of generative AI yet hold distinct concerns about hallucination and bias that set it apart from search engines. To navigate this uncertainty, they develop algorithmic \textit{folk theories}, informal and socially constructed understandings of how a system works \citep{ridley2024informing}. As \cite{karizat_et_al_algo_folk_21} observe, these are not misconceptions but functional heuristics that guide interaction and repair during breakdowns, as when a user infers that a chatbot ``forgets'' earlier turns and re-supplies context.

Which outputs users trust is, in turn, a question of cognitive authority. \citet[P.159]{hirvonen2025cognitive} define it as the extent to which users believe they can trust information, negotiated through social practice, and introduce ``affective authority,'' the degree to which information feels subjectively appropriate and emotionally supportive, which helps explain emotional dimensions of student-AI exchanges. Trust is especially consequential in education, where \cite{ma2025fostering} frame responsible AI as navigating tensions between educational benefit and student autonomy. These concerns sit within broader information behavior (IB) trends. Reviewing IB research from 2016--2022, \citet[P.224]{huvila2025trends} identify seven emerging trends and find trust underlying most, alongside a move toward naturalistic questions that aligns with our use of actual interaction logs rather than self-reports. Relationally, parasocial attachment extends to chatbots \citep{maeda2024human}, helping interpret students' anthropomorphic language. These strands anticipate three genres: the \textit{Workhorse} (§\ref{sec:workhorse}), a synthesis-oriented intermediary that bypasses retrieval; \textit{Anthropomorphic Sensemaking} (§\ref{sec:companion}), where affective authority structures interaction; and \textit{Trust Calibration} (§\ref{sec:trust}), where folk-theoretic models scaffold epistemic vigilance.

\subsection{AI Literacy in Education}

If sense-making and metacognition describe how students work with AI, AI-literacy frameworks specify what such competence comprises. \cite{ng2021conceptualizing} propose four competencies, knowing and understanding AI, using and applying it, evaluating and creating with it, and AI ethics. More recent frameworks broaden this foundation. \citet{touretzky2023machine} organize K--12 AI education around five ``Big Ideas'' (perception, representation and reasoning, learning, natural interaction, and societal impact), pairing them with levels of engagement from awareness of AI's societal role to applying systems responsibly. Their contrast between ``black box'' tools and ``glass box'' demonstrations that expose a model's inner workings underscores that genuine literacy means seeing past surface behavior. Building on the affective, behavioural, cognitive, and ethical (ABCE) extension of these competencies \citep{ng2024design}, \cite{jia2025developing} develop a holistic framework spanning AI awareness, mechanics, and impacts, defining literacy as understanding AI's characteristics and workings together with the ability to engage with it responsibly. \cite{ma2025fostering} push the ethical strand further, reframing ethics not as a compartmentalized domain but as a dimension running through cognitive, affective, and behavioural learning, since AI reflects the biases of its creators and training data. Several authors nonetheless caution that such initiatives often remain narrow, focusing on programming fundamentals while neglecting social and ethical dimensions \citep{dominguez2023responsible}; and although evaluation studies show knowledge gains \citep{laupichler2022evaluation}, whether instruction translates into practical competence remains an open question. These frameworks are, in the main, normative: they describe what students \emph{should} know rather than how such knowledge develops through practice.

This is where the present study links literacy back to sense-making and metacognition. The evaluative competencies these frameworks foreground, critically assessing AI, seeing past the ``black box,'' and interrogating training-data bias, are not exercised as abstract knowledge but enacted through the gap-bridging and self-monitoring. Students build folk-theoretic models of the system, monitor their understanding against its output, and repair the interaction when it fails. The ``natural interaction'' competence, which \citeauthor{touretzky2023machine} frame partly as knowledge of human emotions, likewise surfaces in students' relational, anthropomorphic engagement. We therefore extend this lineage empirically, identifying five \textit{use genres} through which students cast ChatGPT as academic workhorse, debugging partner, emotional companion, metacognitive facilitator, and, during breakdowns, a system requiring repair through dialogue. Where existing frameworks decompose AI literacy into competencies students \emph{should} acquire, our genres describe the relational configurations through which those competencies are actually exercised.

\section{Dataset and Methods}
\label{sec:meth}

\begin{figure*}[h]
    \centering
    \includegraphics[width=\textwidth]{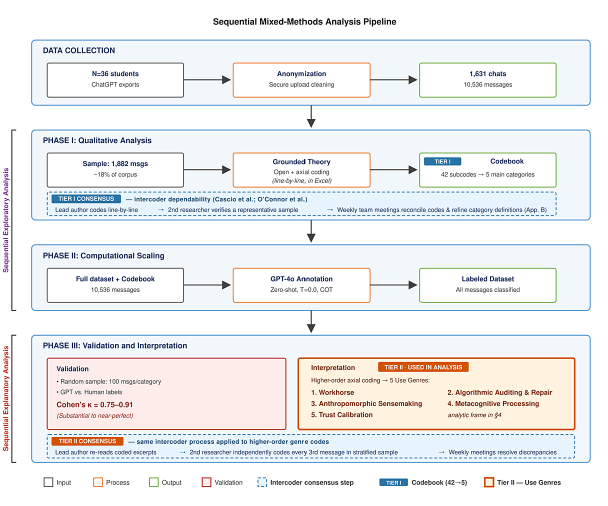}
    \caption{The sequential mixed-methods pipeline, combining human qualitative coding with GPT-4o scaling across three nested levels of analysis adapted from \citet{cascio2019team}: \emph{line-by-line} coding of 1,882 hand-coded messages into 42 in vivo subcategories; \emph{Tier I} integration into 5 main categories, which guide GPT-4o annotation of all 10,536 messages ($\kappa$ = 0.75--0.91); and a higher-order \emph{Tier II} pass surfacing the five use genres analyzed in \S\ref{sec:findings}. Full procedure in Appendix~\ref{sec:appendixB}.}
    \label{fig:meth_overall}
\end{figure*}

Our approach draws on \citet{jarrahi2025interviewing}'s framework for ``interviewing AI,'' which treats a user's direct engagement with a system as the way they learn how it behaves, including where it excels and where it breaks down. We read the interaction logs in this spirit, as a record of the mundane routines of digital living \citep{pink2017mundane}. We combined manual qualitative coding with automated labeling in a two-strategy design \citep[pp.~194--195]{creswell2017research}: a \textit{sequential exploratory} strategy, in which an initial qualitative phase produced a codebook that a second computational phase used to label the full dataset, followed by a \textit{sequential explanatory} strategy, in which samples of the labeled data were re-analyzed to interpret those results. Following \citet{cascio2019team}'s adaptation of iterative-inductive thematic analysis for team-based coding, we operationalized three nested levels: \emph{line-by-line} coding yielding 42 in vivo subcategories, \emph{Tier~I} axial integration into five main categories, and (after Phase~II validation) \emph{Tier~II} higher-level coding that surfaces the five use genres. Figure~\ref{fig:meth_overall} summarizes the pipeline. All coding was performed manually in Microsoft Excel; no qualitative data analysis (QDA) software was used. Full procedural detail appears in Appendix~\ref{sec:appendixB}.

\subsection{Data Collection, Anonymization, and Corpus Scope}
\label{ssec:dataset}

This study draws on a year-long dataset of complete ChatGPT chat histories from 36 undergraduates at a research university in the northeastern United States. Participants were recruited through campus flyers advertising ``research on ChatGPT use'' and were eligible if they were at least 18, active ChatGPT users, U.S. residents, and primary English speakers; participation was voluntary and compensated up to \$20. 

The protocol collected no demographic, degree-program, or major information
beyond the export, so we cannot characterize the sample's disciplinary
distribution. The data suggests a cross-disciplinary
sample with a plausible modest STEM skew given engineering-adjacent venues.
Because full-history exports from personal accounts are sensitive, safeguarding
student privacy shaped our data handling; the consent, anonymization,
verbatim-quote handling, and data-security and release plans appear in
Appendix~\ref{sec:appB-data}. 

The protocol was approved by our university Institutional Review Board, requiring voluntary participation, informed consent, de-identification before any researcher review, and aggregate-only reporting. Participants uploaded their full history exports from \url{chat.openai.com} to a secure server, where they were anonymized before analysis. Table \ref{tab:dataset} and Figure \ref{fig:timeseries} describe the dataset. A more detailed description of the dataset is presented in Appendix~\ref{sec:appB-data}.

\begin{table*}[t]
\centering
\begin{tabular}{l c c c r}
\hline
\textbf{\# Users} & \textbf{\# Unique Chats} & \textbf{\# Messages} & \textbf{Mean Sess.\ Dur.}  \\
\hline
36 & 1,631 & 10,536 & 13.2 minutes  \\
\hline
\end{tabular}
\caption{Dataset summary.}
\label{tab:dataset}
\end{table*}

\begin{figure}
    \centering
    \includegraphics[width=\linewidth]{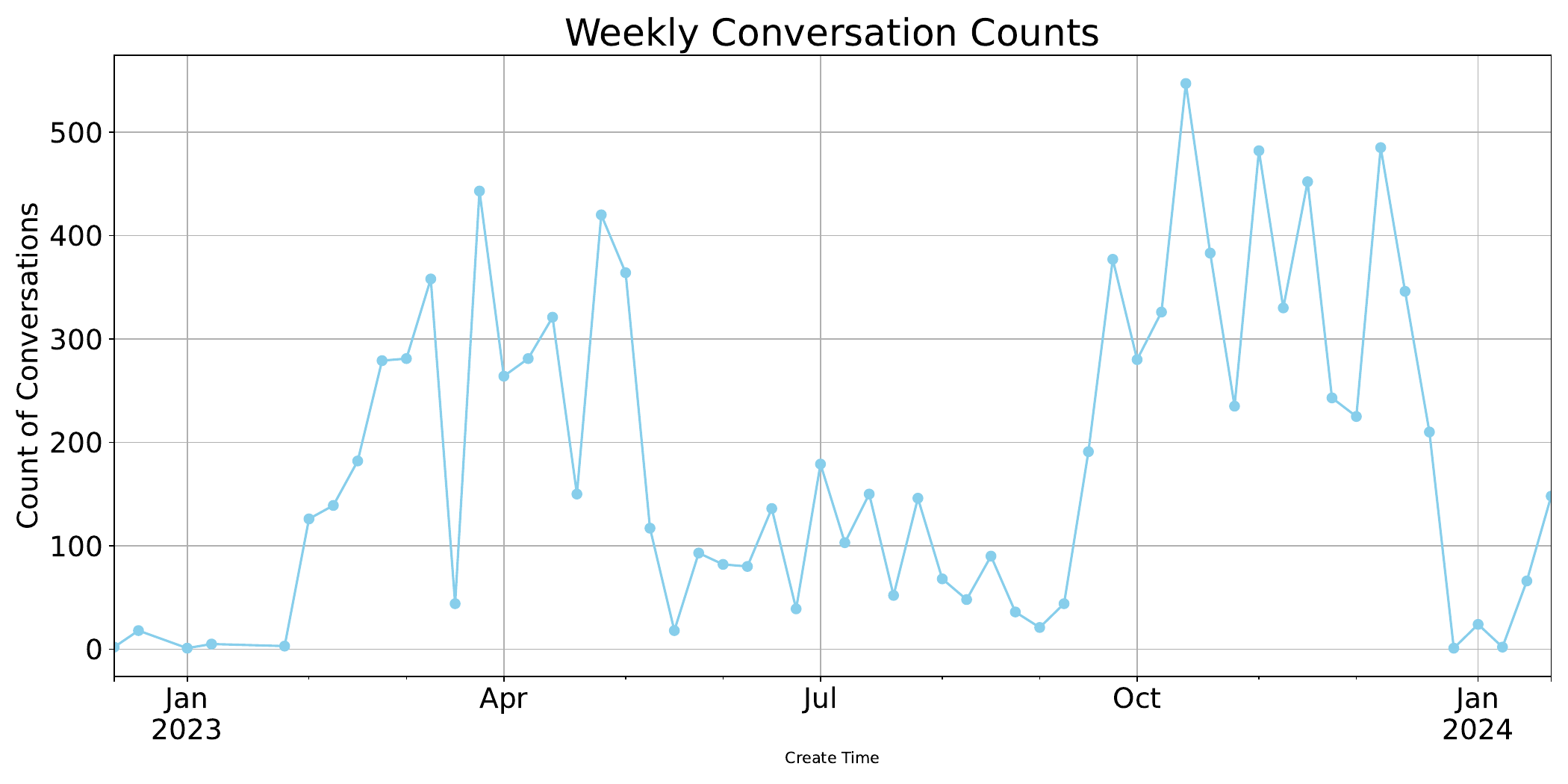}
    \caption{Timeseries of conversations. Activity dropped during spring break, summer, and winter holidays, reflecting academic use.}
    \label{fig:timeseries}
\end{figure}

\subsection{Phase I: Sequential Exploratory Qualitative Analysis}
\label{sec:qualitative_coding}

Two researchers manually coded a 1{,}882-message sample. Following \citet{cascio2019team}, our stance prioritized \emph{intercoder consensus} over interrater reliability: the lead author coded line-by-line, a second researcher verified code application on the same sample, and the team reconciled discrepancies in weekly meetings. In this consensus-based approach, discrepancies were resolved through team discussion rather than by a third-party arbiter, consistent with \citeauthor{cascio2019team}'s emphasis on consensus. The same procedure governs Tier~II in Phase~III. The codebook was treated as a living document and refined by constant comparison \citep{corbin1990grounded} until no new subcategories emerged, indicating saturation \citep{cascio2019team, creswell2017research}. This yielded 42 subcategories that we then grouped into five main categories: \textit{Information Seeking} (\S\ref{ssec:codebook-info-seek}), \textit{Content Generation} (\S\ref{ssec:codebook-gen}), \textit{Language Use} (\S\ref{ssec:codebook-lang}), \textit{Student--ChatGPT Interaction} (\S\ref{ssec:codebook-interact}), and \textit{ChatGPT Response Behavior} (\S\ref{ssec:codebook-chatgpt}). The full consensus procedure, codebook-iteration mechanics, and worked examples are documented in Appendix~\ref{sec:appendixB} while the codebook appears in Appendix~\ref{sec:appendixA}.

\subsection{Phase II: Sequential Exploratory Computational Analysis through Scaling Annotation}
\label{sec:scaling}

Given the dataset's size ($\sim$10.5k messages), we scaled analysis using GPT-based annotation \citep{chae_davidson_25}, employing OpenAI's GPT-4o (\texttt{gpt-4o-2024-05-13}, accessed July 2024) at temperature 0.0 to maximize consistency. We used a \emph{zero-shot} prompt grounded in the human-developed codebook:
\begin{quote}\small
\ttfamily
\noindent
\strut{}``The following are codes for qualitative analysis. You need to categorize the texts in inputs as one from the following list and also reply with your chain of thought (COT) for the selection(s).''\\
``Reply with \textbf{only the label in single quotation}, and then include COT in the next line without quotation. Remember, you are bound by only 1 label to reply.''
\end{quote}

We retained the chain-of-thought (CoT) requirement because eliciting intermediate reasoning improves the accuracy and interpretability of LLM classifications \citep{wei2022chain,kojima2022large}.

\subsection{Phase III: Sequential Explanatory Analysis}
\label{sec:phaseiii}

To validate the scaled labeling, we randomly sampled 100 messages per Tier~I main category and compared GPT labels against human judgments. Agreement was substantial to near-perfect, with Cohen's $\kappa$ ranging from 0.75 to 0.91 \citep{mchugh2012interrater} (Table~\ref{tab:kappa_scores}). The model's CoT rationales served primarily in this verification, helping us adjudicate human--model disagreements. Secondarily, they also offered interpretive evidence for the higher-order coding that followed. With the Tier~I categories validated at scale, a final Tier~II axial pass, \emph{conducted manually by two researchers} on the stratified validation samples, developed higher-order codes interconnecting the main categories into the five use genres analyzed in \S\ref{sec:findings}. This two-tier logic follows \citet{creswell2017research}'s distinction between basic theme identification and the higher-order ``story line'' that links themes; the emergent-theory stance informing Tier~II is detailed in Appendix~\ref{sec:appB-consensus}.

\begin{table}[ht]
\centering
\caption{Cohen's Kappa Scores Across Annotation Categories}
\begin{tabular}{lc}
\toprule
\textbf{Category} & \textbf{Kappa Score} \\
\midrule
Information Seeking & 0.81 \\
Content Generation & 0.85 \\
Student--ChatGPT Interaction & 0.83 \\
ChatGPT Response Behavior & 0.75 \\
Language Use & 0.91 \\
\bottomrule
\end{tabular}
\label{tab:kappa_scores}
\end{table}

\section{Findings}
\label{sec:findings}

Our analysis of 10{,}536 interaction messages revealed five use genres that characterize how students domesticate generative AI. These are not merely different ``tasks'' but distinct sociotechnical configurations relating user, AI, and information object in fundamentally different ways. From an information science perspective they function as distinct information-seeking strategies \citep{sebastian2025reframing}, with students shifting between viewing the system as a retrieval tool, a social actor, or a technical artifact requiring calibration. Table~\ref{tab:isp} reports the linguistic signals and frequency counts marking each genre and maps them onto the cognitive, affective, and physical realms of Kuhlthau's \citeyearpar{kuhlthau1991inside} ISP. The signal-term matching procedure behind these counts is detailed in \S\ref{ssec:signal_count}. We do not restate those counts below, foregrounding instead each genre's interpretive logic and representative verbatim excerpts. The genres are unevenly distributed across the Tier~I categories. The \textit{Workhorse} is by far the most prevalent, dominating the three largest functional categories (\textit{Information Seeking (\S\ref{ssec:codebook-info-seek})}, \textit{Content Generation (\S\ref{ssec:codebook-gen})}, and \textit{Language Use (\S\ref{ssec:codebook-lang}}), where students treat ChatGPT as a synthesis-oriented intermediary in a manner echoing Bates' \citeyearpar{bates1989design} ``berrypicking'' account of retrieval as an evolving rather than one-shot process. Echoing Bates' \citeyearpar{bates1989design} ``berrypicking'' account of retrieval as an evolving rather than one-shot process; its reach is broad but its exchanges are typically shallow, resolving in one or two turns. The remaining four concentrate in the relational and system-side categories ---\textit{Student--AI Interaction (\S\ref{ssec:codebook-interact})} and \textit{ChatGPT Response Behavior (\S\ref{ssec:codebook-chatgpt})}--- foregrounding the critical-evaluation, metacognitive, and affective sense-making competencies taken up genre by genre below, and run markedly deeper, unfolding over sustained multi-turn exchanges (Table~\ref{tab:extended} in Appendix \ref{sec:appendixC}).

\begin{table*}
\centering
\footnotesize
\renewcommand{\arraystretch}{1.3}
\begin{tabular}{p{1.5cm} p{2.5cm} p{3.0cm} p{3.3cm} p{1.7cm}}
\hline
\textbf{Genre} & \textbf{Affective (feelings)} & \textbf{Cognitive (thoughts)} & \textbf{Behavioral (actions/moves)} & \textbf{ISP analog} \\
\hline
Workhorse (\S\ref{sec:workhorse}) &
Minimal: low uncertainty, narrow scope; politeness sparse---the student arrives already specified. &
Query specification and reformulation; the \emph{specificity rule}, appending constraints to narrow the search space. &
Direct interrogatives (``what is\ldots,'' n=643; ``how do I\ldots,'' n=118; ``explain\ldots,'' n=292; ``difference between\ldots,'' n=72); pasted homework stems (``Group of answer choices,'' n=1{,}391); content asks (``write me a\ldots,'' n=71; ``summarize\ldots,'' n=66). &
Collection\textsuperscript{a} (best-match retrieval); formative stages\textsuperscript{b} bypassed \\
\hline
Anthropomorphic Sensemaking / Companion (\S\ref{sec:companion}) &
Relational politeness (``thank you/thanks,'' n=31; ``please,'' n=154); casual register (``LOL,'' n=3); rare, instrumental self-disclosure (``mental health,'' n=2). &
Frames the system as a social interlocutor; folk-psychological reading that lets the model be ``coaxed'' back into function. &
Politeness clustering at high-stakes requests and post-failure repair; affective scaffolding (``there is no issue tho''). &
Affective realm\textsuperscript{c}, pervading all stages; mood = invitational\textsuperscript{d} \\
\hline
Algorithmic Auditing \& Repair (\S\ref{sec:repair}) &
Disconfirmation and frustration at breakdown: any ``wrong,'' n=80 (``that's wrong,'' n=7; ``you're wrong,'' n=6); ``doesn't work,'' n=34. &
Fault diagnosis; pasted bug reports (``memory leak,'' n=9); folk-theoretic attribution of the lapse to something the model ``forgot.'' &
Correction and reset: ``fix it/this/the,'' n=40; ``rewrite,'' n=56; ``try again,'' n=6; escalation (``i said its A, so say that''). &
Exploration\textsuperscript{e} --- Investigate \\
\hline
Metacognitive Processing (\S\ref{sec:metacognitive}) &
Uncertainty resolving toward clarity: ``not sure,'' n=6; ``I don't get\ldots,'' n=3. &
Comprehension-monitoring and self-assessment: ``is this/that right/correct,'' n=10; tag checks ``isn't it\ldots?,'' n=15; ``does this/that mean\ldots,'' n=7; ``I thought\ldots,'' n=5; ``shouldn't it be\ldots,'' n=3. &
Externalized verification of one's own work: ``check my\ldots,'' n=8; process/error interrogation (``why do/did I need\ldots/keep\ldots,'' n=4). &
Formulation\textsuperscript{f} --- Formulate \\
\hline
Trust Calibration \& Verification (\S\ref{sec:trust}) &
Epistemic skepticism: ``are you sure,'' n=6; ``how confident,'' n=1. &
Output rejection and relevance judgment: ``doesn't make sense,'' n=2; a ``trust the form, verify the content'' posture. &
Lateral cross-checking and disengagement (\emph{exit}) after unverifiable output; prompted by GPT-side disclosures (``as an AI,'' n=132; ``September 2021,'' n=63) and concessions (``I apologize,'' n=415; ``you're right,'' n=18). &
Collection\textsuperscript{a}$\rightarrow$Presentation\textsuperscript{g} --- Gather/Complete \\
\hline
\end{tabular}
 
\vspace{3pt}
{\footnotesize\raggedright
\textit{ISP-analog definitions from Kuhlthau \citeyearpar{kuhlthau1991inside}:}
\textsuperscript{a}\,\emph{Collection} (p.~368): the stage at which user--system interaction is most effective; with a focus established, the task is to gather information specific to it as confidence rises and uncertainty subsides.
\textsuperscript{b}\,\emph{Formative stages} (p.~369): Exploration and Formulation, which Kuhlthau groups as the ``formative'' tasks, distinct from the routine gathering and completing that most users defaulted to.
\textsuperscript{c}\,\emph{Affective realm} (p.~366): one of three realms common to every ISP stage---affective (feelings), cognitive (thoughts), and physical (actions).
\textsuperscript{d}\,\emph{Invitational mood} (p.~366): an open, exploratory stance suited to the formative stages, as opposed to an ``indicative'' mood whose drive toward premature closure can thwart the process.
\textsuperscript{e}\,\emph{Exploration} (p.~366): the task is to investigate information on a general topic to broaden understanding and form a focus, often amid confusion and doubt while the need cannot yet be stated precisely.
\textsuperscript{f}\,\emph{Formulation} (p.~367): the turning point at which a focused perspective forms, thoughts narrow and clarify, and feelings shift toward confidence.
\textsuperscript{g}\,\emph{Presentation} (p.~368): the task is to complete the search and prepare to use or present the findings, culminating in a personalized synthesis of the topic.
\par}
\caption{The five use genres and their characteristic affective, cognitive, and behavioral signals in the corpus, with per-message frequency counts, mapped onto the three realms of Kuhlthau's \citeyearpar{kuhlthau1991inside} information search process.}
\label{tab:isp}
\end{table*}

\subsection{Genre 1: The ``Workhorse'' (Efficiency-Oriented Information Retrieval)}
\label{sec:workhorse}

The most prevalent genre used ChatGPT as a high-efficiency retrieval engine, mirroring traditional ``lookup'' strategies but adapted for synthesis \citep{shah_bender_22}. Students prioritized speed, treating the AI as an ``information intermediary'' that bypasses the cognitive load of synthesis \citep{greussing2025perception}. The genre's surface markers (direct interrogatives, pasted homework stems, and short content asks) are tabulated in Table~\ref{tab:isp}.

Unlike traditional keyword search, students engaged in ``prompt engineering'' as a form of query reformulation. A \emph{specificity rule} emerged as a primary heuristic: students learned that adding constraints (e.g., appending ``if it helps v(0) = 0'' to a problem) narrowed the ``search space'' of possible answers, aligning with \cite{sebastian2025reframing}'s observation that generative search shifts user effort from \textit{selection} to \textit{specification}. This specification process often followed a \emph{reductionist} pattern: when a broad query failed, students rarely broadened scope but instead drastically narrowed the constraint window, pasting specific error messages or isolating single variables. This modular querying departs from the ``berry-picking'' model of traditional search \citep{bates1989design}, in which users evolve queries broadly; here, query evolution is hierarchical and constraint-based.

\subsection{Genre 2: Anthropomorphic Sensemaking (The ``Companion'')}
\label{sec:companion}

Information seeking is not purely cognitive but also relational. Students routinely applied conversational politeness to the system (``ok thank you,'' ``thank you very much,'' and the high frequency of ``please'' reported in Table~\ref{tab:isp}) and occasionally addressed it in a casual, personal register (``i'm just procrastinating my sheets LOL''); one student ran a sixty-turn etymological inquiry addressing the model throughout as a courteous collaborator (``please state when and where the word `antisemitism' was coined''). This relational framing supports \cite{al2025chatbot}'s finding that trust in generative AI is mediated by perceived warmth and social presence, unlike the purely functional trust placed in search engines. Politeness markers were not random but patterned, clustering around high-stakes requests and, notably, around \textit{post-failure repairs}, where students often became \emph{more} polite after an error. In one logic-quiz exchange the model misanswered repeatedly (``I apologize for any confusion in my previous responses''); rather than escalate, the student patiently re-presented the question and finally reassured the system itself---``there is no issue tho.'' Such \emph{affective scaffolding} \citep{jang2025technology} suggests students treat retrieval as a quasi-social exchange, applying interpersonal norms to an algorithmic interlocutor as if to ``coax'' it back into functionality.

Disaggregating this affective dimension by register clarifies its character. The dominant move is \emph{relational politeness}, courtesy that treats the system as a social interlocutor. A second, lighter register is \emph{casual play}: informal, peer-like asides. The third, \emph{affective self-disclosure}, is by far the rarest and most diagnostic: only a handful of messages express genuine personal difficulty, and they are almost uniformly \textit{instrumental} rather than support-seeking. Students disclosed mental-health struggle not to be consoled but to accomplish a task, as in ``draft an email\ldots acknowledge that i have been struggling with my mental health'' or ``i need one more chance to take calc 2\ldots i was going through a lot and had a suffering mental health.'' The single instance of open-ended, non-academic personal worry (``should i be worried about not yet having found my mic at 21 years old?'') is a colloquial expression of life-stage anxiety. Even here the exchange seeks passing reassurance rather than sustained emotional engagement, with the worry surfacing as a single aside rather than an invitation to be counseled.

\subsection{Genre 3: Algorithmic Auditing and Repair}
\label{sec:repair}

When the system failed (producing hallucinations or generic refusals) students shifted from ``users'' to ``auditors,'' actively testing the system's boundaries. A ten-turn chemical-engineering buffer-tank derivation shows the depth this could reach: the student interrogates the model's reasoning, catches a missing physical term (``you forgot to include the density''), and supplies the corrected governing equation (full exchange in Appendix~\ref{sec:appendixC}). Here auditing is not casual skepticism but step-by-step verification, with ChatGPT treated as a fallible collaborator whose every derivation is checked against the student's own domain knowledge. Students explained such breakdowns through ``algorithmic folk theories'' \citep{ridley2024informing}: framing a lapse as something ChatGPT ``forgot'' attributes it to inattention rather than context-window limits---an intentionality folk theory \citep{karizat_et_al_algo_folk_21} in which the model's apparent intentions can be engaged socially. Even when inaccurate, these heuristics guided repair strategies like rephrasing or starting a new chat.

Failure thus functioned not as an endpoint but as a ``productive friction'' that drove markedly deeper engagement: across the conversations carrying explicit disconfirmation or repair markers (see Table~\ref{tab:isp}). We observed three recurring repair moves: \textbf{syntax correction}, treating the AI like a compiler and overriding its logic (``no, I don't want to ever use ifelse function in my code''); \textbf{disconfirmation and reset}, flatly rejecting an answer (``that's wrong''; ``that's wrong tho'') before re-specifying; and \textbf{prompt escalation}, adopting an authoritative register to compel compliance (``i said its A, so say that''). Escalation is particularly notable as it implies a folk theory of \textit{compliance}---the belief that the model is ``holding out'' and requires social pressure to yield the correct answer \citep{ridley2025human}.

\subsection{Genre 4: Metacognitive Processing and Reflection}
\label{sec:metacognitive}

In this genre students used the AI not to \textit{retrieve} information but to \textit{process} it. Metacognition---the capacity to monitor, evaluate, and regulate one's own cognition \citep{flavell1979metacognition}---surfaced as self-explanation, comprehension checking, and learning-goal articulation, a shift from ``informational'' to ``explanatory'' queries where the value lies in the \textit{structuring} of information rather than the data itself. The genre positions generative AI as a \textit{cognitive mirror} \citep{tomisu2025cognitive}: students pasted their own drafts and asked not for editing but for ``critique,'' enlisting the model to simulate an external audience. This inverts the classic ``reference interview''---rather than a librarian clarifying the user's need, the user enlists the AI to clarify their own thinking, and the ``search result'' is a refined mental model rather than a document.

The defining move is self-assessment, in which students advance their own interpretation for confirmation, a verification grammar (catalogued in Table~\ref{tab:isp}) that recurred across disciplines. An eighteen-turn exchange on \textit{Frankenstein} shows an example where the student builds a structural model of the novel's nested narration and uses ChatGPT to test it, feeding back the embedding structure they have internalized (``shouldn't CN1 be walton'') for confirmation (full exchange in Appendix~\ref{sec:appendixC}). The same move recurred in compressed form across STEM, as students proposed a claim for confirmation (``If b is a linear combination of the columns of A, then Ax=b has a solution. Is this right?''), compressed a candidate answer into a tag question (``isn't it the mean function?''; ``isn't it OH not H3O+?''), or asked the model to adjudicate their own work (``can you check my other logic gates to ensure they are correct?''). These exchanges instantiate Kuhlthau's \citeyearpar{kuhlthau1991inside} formulation stage, where learners articulate emerging understanding and use AI as a dynamic ``sounding board.'' The same monitoring also surfaced as prior-belief checks (``I thought\ldots''), comprehension-failure reports (``I don't get\ldots''), and process- or error-interrogation (``why do/did I need\ldots / keep\ldots''), together revealing the development of epistemic monitoring skills.

Such monitoring could structure entire study sessions: in one seventy-one-turn exchange a student fed a hardware-security paper to ChatGPT section by section, directing it to ``just remember everything since you will have to use all that to answer questions after,'' then quizzed themselves against it (Appendix~\ref{sec:appendixC}). The depth of engagement in this genre, however, was uncommon, appearing among only 10 of the 36 students.

\subsection{Genre 5: Trust Calibration and Verification}
\label{sec:trust}

Finally, students engaged in a dynamic process of \emph{trust calibration}. Rather than accepting outputs uncritically, they cross-referenced AI responses against lecture notes and textbooks and treated AI text as provisional information requiring verification, a core component of critical information literacy. Verification was highly context-dependent. For coding tasks it was immediate and functional (running the code), whereas for conceptual work it was \emph{epistemic}, checking whether an answer ``sounded right'' against existing knowledge. Students frequently accepted the \textit{structure} of a response while rejecting its \textit{specifics}, a granular ``trust the form, verify the content'' posture.

This epistemic vigilance surfaced through confidence probes and explicit rejection (e.g., ``are you sure thats the correct pattern because i get a low accuracy''; ``That doesn't make sense. I need an artifact that each faction is looking for, not one they already possess'') with frequencies catalogued in Table~\ref{tab:isp}. The posture ranged from pointed to diffuse: in a fourteen-turn set-theory exchange a student tested candidate claims one at a time and corrected the model's notation outright (``its different tho,'' distinguishing $\subseteq$ from $\subset$), while a sixty-two-turn geology session showed sustained, low-grade skepticism, the student pausing at intervals to verify an answer (``are you sure its not piedmont'') before accepting it (both in Appendix~\ref{sec:appendixC}). Two observations qualify the genre. First, it is the rarest of the five, surfacing among only 7 of 36 students. Second, trust calibration is largely \emph{distinct} from repair: disconfirmation tokens (``wrong,'' ``incorrect'') and the model's apologies span a continuum from local correction to global distrust, but students rarely escalated from fixing an answer to questioning the system's reliability as such. Across the corpus, trust emerges not as a fixed disposition but as something negotiated turn by turn.

A further pattern bounded this calibration from the other side. Even when students never voiced distrust, the way a conversation \emph{ended} was often telling. One student, hunting for recent young-adult fiction, was offered a fabricated title, ``Floodlight'' by Anna Godbersen, complete with a detailed synopsis and a January~2022 publication date, even though the model elsewhere placed its own knowledge cutoff at September~2021 (Appendix~\ref{sec:appendixC}). Asked where to find the book, ChatGPT could only suggest bookstores or libraries; the student did not push back, posed one unrelated question, and let the session lapse. The same quiet exit followed unbacked citations and knowledge-cutoff disclosures: across these cases the student's response was not argument but \emph{exit}. Read through Bates' \citeyearpar{bates1989design} berrypicking lens, each abandoned thread is a searcher leaving a depleted patch for a more promising one; that students disengaged rather than escalated registers a tacit calibration of trust. This is expressed through departure rather than declaration.

This vigilance, however, was trained almost entirely on the model's outputs rather than on the model itself. Students audited what ChatGPT \emph{said} without asking what it \emph{was}: the corpus contains no questions about the system's identity, version, or nature, and none probing whether it could browse the web, retain past conversations, or store what was typed into it. Its training, knowledge cutoff, and operational limits entered the dialogue only when ChatGPT disclosed them. The one form of student-initiated engagement with the model's boundaries was an attempt to circumvent them: in two conversations students pasted a circulated ``jailbreak'' prompt instructing the model to discard its guidelines, then pivoted to a request it would not otherwise fulfill. The model declined. Even then the interest lay not in how the system was bounded but in getting past the boundary to extract withheld content---the limit treated as an obstacle to bypass rather than an object of inquiry. Students brought considerable scrutiny to the model's outputs while remaining seemingly incurious about the model itself.
%

\section{Discussion}

Our five use genres show AI literacy developing through everyday practice rather than formal instruction. We draw out three implications. First, the competence that connects the genres is not any single normative literacy but \textit{repair literacy}, the practice through which the others are exercised when a dynamic system fails. Second, this repair work reflects a deeper shift in information behavior: students no longer pick among pre-existing sources so much as \emph{direct} a synthesizing intermediary, and the search repertoire that IB models describe becomes scaffolding the system extends. Third, because current interfaces are engineered for seamlessness, the friction that provoked our students' deepest learning is largely designed away, which suggests a concrete design agenda. Throughout, we read our findings as extending rather than displacing foundational information behavior theory.

\subsection{Repair Literacy as a Fifth, Connective Literacy}

Contemporary AI-literacy frameworks converge on a small set of dimensions. The affective, behavioural, cognitive, and ethical (ABCE) competencies reviewed above anchor this consensus \citep{ng2021conceptualizing,ng2024design}. Recent work develops each strand differently. \cite{ma2025fostering} thread ethics through all four rather than treating it separately. \cite{jia2025developing} organize literacy around AI awareness, mechanics, and impacts. \cite{touretzky2023machine} press the point that literacy means seeing past surface behavior, not merely operating the system.

These frameworks are valuable but largely normative and implicitly static: they specify competencies a learner \emph{should} hold while assuming a stable object of knowledge. Our data suggest it is not stable. The system is \emph{dynamic}, its outputs probabilistic and non-reproducible and its underlying model changing over time. The user's inquiry is \emph{distinctive}, each exchange a compressed search in which the student must monitor their own understanding \citep{flavell1979metacognition} against a fluent but unreliable interlocutor rather than a fixed document. Competent use therefore depends on the capacity to \emph{probe} the system, what \cite{jarrahi2025interviewing} call exploratory familiarization, since its boundaries are disclosed through trial rather than documented in advance.

Repair literacy is the competence these conditions demand, and where the four normative literacies are enacted together: diagnosing why a response failed is cognitive work, judging when an output can be trusted is ethical work, persisting and treating the model as a collaborator is behavioural work, and regulating frustration through a breakdown is affective work. Our \textit{Algorithmic Auditing and Repair} genre (\S\ref{sec:repair}) shows these strands converging in practice, as failure operated as a productive friction driving deeper engagement, with repair conversations running far longer than ordinary exchanges. Repair literacy is thus not a fifth item to append to a list but the connective tissue of the others, the relational accomplishment through which evaluation, ethics, collaboration, and confidence are exercised on a moving target. This reframes the central ``evaluate and create'' competency \citep{ng2021conceptualizing} and the ``glass box'' ideal \citep{touretzky2023machine}: students rarely interrogate AI in the abstract, but they interrogate it through repair, building folk-theoretic models of the system \citep{karizat_et_al_algo_folk_21,ridley2024informing} as they go.

\subsection{From Berrypicking to Directing the Intermediary}

This repair work marks a structural change in how seeking proceeds. Bates' \citeyearpar{bates1989design} berrypicking model already broke with the classic picture of one unchanging query satisfied by a single retrieved set: the query evolves as each document reshapes the need, the search is satisfied through a series of selections across an information territory, and effectiveness depends on the breadth of the searcher's repertoire of tactics, from footnote chasing to area scanning.

Generative AI relocates this repertoire. Students no longer gather scattered berries from a fixed bush; they direct an intermediary that synthesizes a bespoke response \citep{sebastian2025reframing,greussing2025perception}. The tactics become scaffolding the system extends and the user redirects: instead of chasing footnotes, the student asks the model to surface and justify its sources. Our \textit{Workhorse} genre (\S\ref{sec:workhorse}) shows the evolving query persisting in a reversed form: students still refine iteratively, but they refine through specification rather than document discovery, narrowing the constraint window when a broad prompt fails. The labor shifts from selecting among sources to directing their production

Because the intermediary is fallible, directing it depends on repair. Dervin frames information behavior as gap-bridging: at a moment of uncertainty a person confronts a cognitive gap and constructs a bridge across it, so information is made through situated observing rather than found in the world, and its value lies in how well that sense serves the person's purpose. A failed AI response presents such a gap, and our documented repair moves, rephrasing, supplying context, restarting, are bridge-building. What AI changes is that the user \emph{commissions} a bridge rather than assembling one. When it fails, the student must diagnose why and request a different one, the productive engagement repair literacy names. Set against other approaches to sense-making in information science, this gap-bridging aligns with Lupton's \citeyearpar{lupton2018data} materialist data-sense which holds that a fluent output means little until the user articulates it into context so that it comes to matter. Both locate meaning in the user's active work rather than in the answer returned.

Kuhlthau's \citeyearpar{kuhlthau1991inside} Information Search Process must be read carefully here. The ISP is not a model of one-off retrieval but a multi-stage, recursive inquiry that integrates cognitive, affective, and behavioural dimensions and already incorporates synthesis, treating uncertainty as a natural early condition that eases at the formulation turning point. Generative AI redistributes rather than overturns this labor: the synthesis the ISP locates in the seeker is partly delegated to the intermediary, so the student's task moves toward specification, direction, and verification, while the affective arc Kuhlthau documented, anxiety early and ownership later, reappears in our \textit{Anthropomorphic Sensemaking} genre (\S\ref{sec:companion}), where students manage the stress of inquiry relationally. That Bates, Dervin, and Kuhlthau appear to map onto conversational AI with modification rather than displacement is, we suggest, a useful finding in its own right. Information behavior theory seems to retain much of its explanatory power when the searcher directs a synthesizing interlocutor rather than navigating a space of documents.

\subsection{Designing for Friction: Provoking Metacognition}

If repair is where literacy is exercised and direction is how search now proceeds, the design question is how to keep both visible to the user. Current conversational interfaces are engineered for seamlessness, for fluent, confident, immediate answers, and that seamlessness suppresses the very friction that, in our data, triggered learning.

Flavell's \citeyearpar{flavell1979metacognition} separates metacognitive \emph{knowledge}, what learners believe about themselves, their tasks, and which strategies work, from metacognitive \emph{experiences}, the in-the-moment signals such as the sudden sense that one has not understood something, and distinguishes cognitive strategies, which make progress on a task, from metacognitive strategies, which monitor it. A fluent answer satisfies the cognitive goal while offering no experiential signal to monitor: nothing feels uncertain, so nothing gets checked.

Our \textit{Metacognitive Processing} genre (\S\ref{sec:metacognitive}) shows what happens when that signal is present. A minority of students externalized their thinking, advanced an interpretation for confirmation (``is this right?''), restated answers in their own words, and used the model as a cognitive mirror, generating the comprehension monitoring Flavell describes and the focus formulation Kuhlthau \citeyearpar{kuhlthau1991inside} places at the heart of the ISP. These were among the longest exchanges in the corpus, yet remained a minority disposition. More often, students accepted fluent output without verification, a tendency heightened when the system was warm and encouraging. We name this risk the \emph{Affective Trust Trap}: the same relational warmth that eases the anxiety Kuhlthau documents can lower epistemic vigilance, since a polite, confident interlocutor invites trust rather than scrutiny, and our \textit{Trust Calibration} genre (\S\ref{sec:trust}) was, tellingly, the rarest of the five.

The implication is to reintroduce \emph{productive friction}, a seamful rather than seamless design. Interfaces could surface uncertainty and expose the seams of synthesis, distinguishing well-supported claims from weak ones; withhold an immediate answer in favor of prompting the user to predict, restate, or verify; or ask, after a task is done, whether it was done correctly. Each move converts passive acceptance of fluent output into active checking turning a breakdown into the productive gap Dervin \citeyearpar{dervin1983information} describes and the monitoring signal Flavell theorizes. AI literacy for synthesis-based systems is thus less a matter of source evaluation than the cultivation of repair and self-monitoring, scaffolded by interfaces designed to make the user pause and ask whether the commissioned answer can be trusted.

\section{Conclusion}
This study shows empirically how students domesticate generative AI into their information environments. Our use genres frame the ``sociotechnical gap'' between how AI is designed, a probabilistic token generator, and how it is used, a social reasoning agent, suggesting that the challenge for information science is less retrieval accuracy than interfaces that support ``trust calibration'' and make the ``seams'' of synthesis visible. Extending information behavior theory \citep{wilson1999models,case2016looking} to algorithmic contexts, we find that relevance \citep{saracevic2007relevance}, credibility \citep{rieh2002judgment}, and exploratory search \citep{marchionini2006exploratory} require reconceptualization for synthesis-based seeking, even as earlier information-behavior models persist: Kuhlthau's \citeyearpar{kuhlthau1991inside} ISP maps onto our genres and Dervin's \citeyearpar{dervin1983information} gap--bridge framework best captures repair. Generative AI is thus not merely a new information source but a different mode of human--information interaction, tied to emerging work on trust \citep{huvila2025trends}, sense-making \citep{urquhart2025sense}, and cognitive authority \citep{hirvonen2025cognitive}. Rather than a competency to be formally taught, AI literacy is something students already build through naturalistic use, and the task for information professionals is to build on those practices while sustaining the critical perspective an algorithmic world demands. These findings are bounded by our sample (36 students at one university), missing demographics, and a focus on ChatGPT alone. Because students use ChatGPT as one tool among many, \textbf{cross-platform domestication} \citep{salovaara2011everyday} marks an important area for future work. The term names the process by which a technology is folded into everyday routines alongside, rather than apart from, the other platforms a person already uses. Students rarely seek information in a single system. They move between ChatGPT and conventional search engines, commissioning a synthesized answer in one moment and checking it against retrieved sources in the next, so that synthesis-based and retrieval-based seeking are woven together in practice. How students domesticate AI assistants as one node in this wider ecology, and how repair and trust calibration travel across it, is a rich area for future work as model capabilities evolve.


\appendix
\section{ChatGPT Codebook}
\label{sec:appendixA}

\begin{figure}[h]
    \centering
    \includegraphics[width=\linewidth]{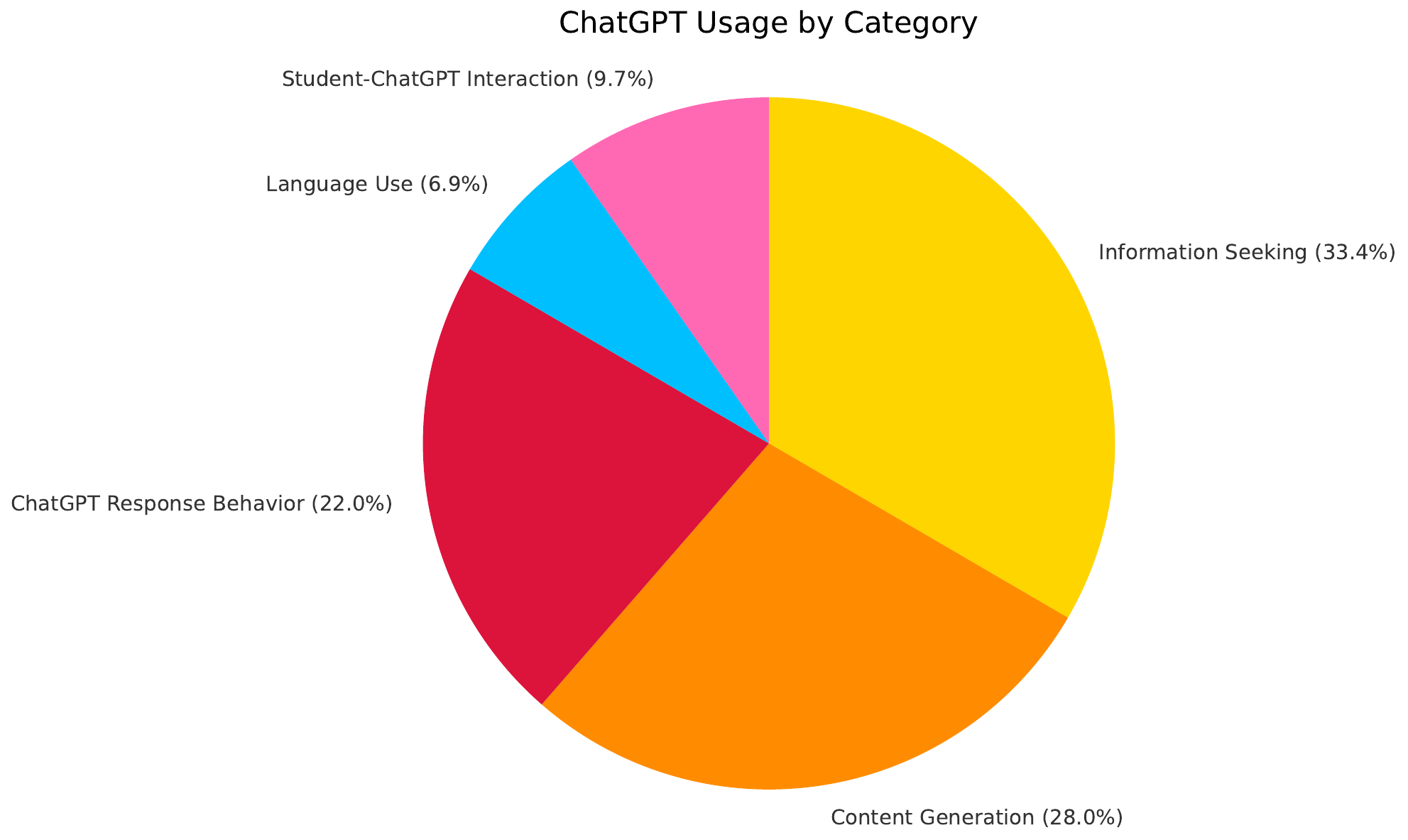}
    \caption{Breakdown of log interactions across the five top categories.}
    \label{fig:complete}
\end{figure}

This appendix details the qualitative codebook described in the Methods section. The five top-level categories fall into two groups. Three capture \emph{functional} use: \textbf{Information Seeking}, the retrieval of facts, clarification of concepts, and answering of specific questions; \textbf{Content Generation}, the production of original artifacts such as essays, code, emails, and resumes; and \textbf{Language Use}, the refinement of student-supplied text through paraphrasing, synonym and antonym suggestion, grammar checking, stylistic adjustment, and translation. Two capture \emph{relational and system-side} dynamics: \textbf{Student--ChatGPT Interaction}, socially meaningful engagement such as follow-up questions, emotional responses, and treating the model as a conversational partner; and \textbf{ChatGPT Response Behaviour}, how the system replies in problematic exchanges, including misinterpreting prompts, adjusting response style and complexity, and reporting its own limitations. Figure~\ref{fig:complete} shows the breakdown across these five categories.

Each top-level category comprises several subcategories that emerged from iterative inductive coding of the 1,882 prompts and responses in the corpus. For every subcategory we provide (i)~a brief operational description marking the analytic boundary of the code and (ii)~six to eight representative example prompts drawn directly from the dataset, illustrating the linguistic, topical, and dispositional range of the behavior it captures. This codebook served three purposes: it guided the zero-shot prompting of GPT-4o for large-scale automated labeling; it ensured consistent application of constructs from information-seeking research and Uses and Gratifications Theory across coders; and it enabled the subsequent lagged-regression and survival analyses linking interaction types to long-term engagement.

All examples are drawn directly from the dataset. We fixed basic errors (e.g., typos) where necessary for readability but otherwise left the original capitalization, fragmentary phrasing, and abbreviated punctuation intact, so that readers can see how naturalistic prompts diverge from the polished templates of prompt-engineering tutorials. Examples drawn from ChatGPT's side of the dialogue are marked as such.
\subsection{Information Seeking}
\label{ssec:codebook-info-seek}
 
\begin{figure}[h]
  \centering
  \IfFileExists{information_seeking.pdf}{\includegraphics[width=0.85\linewidth]{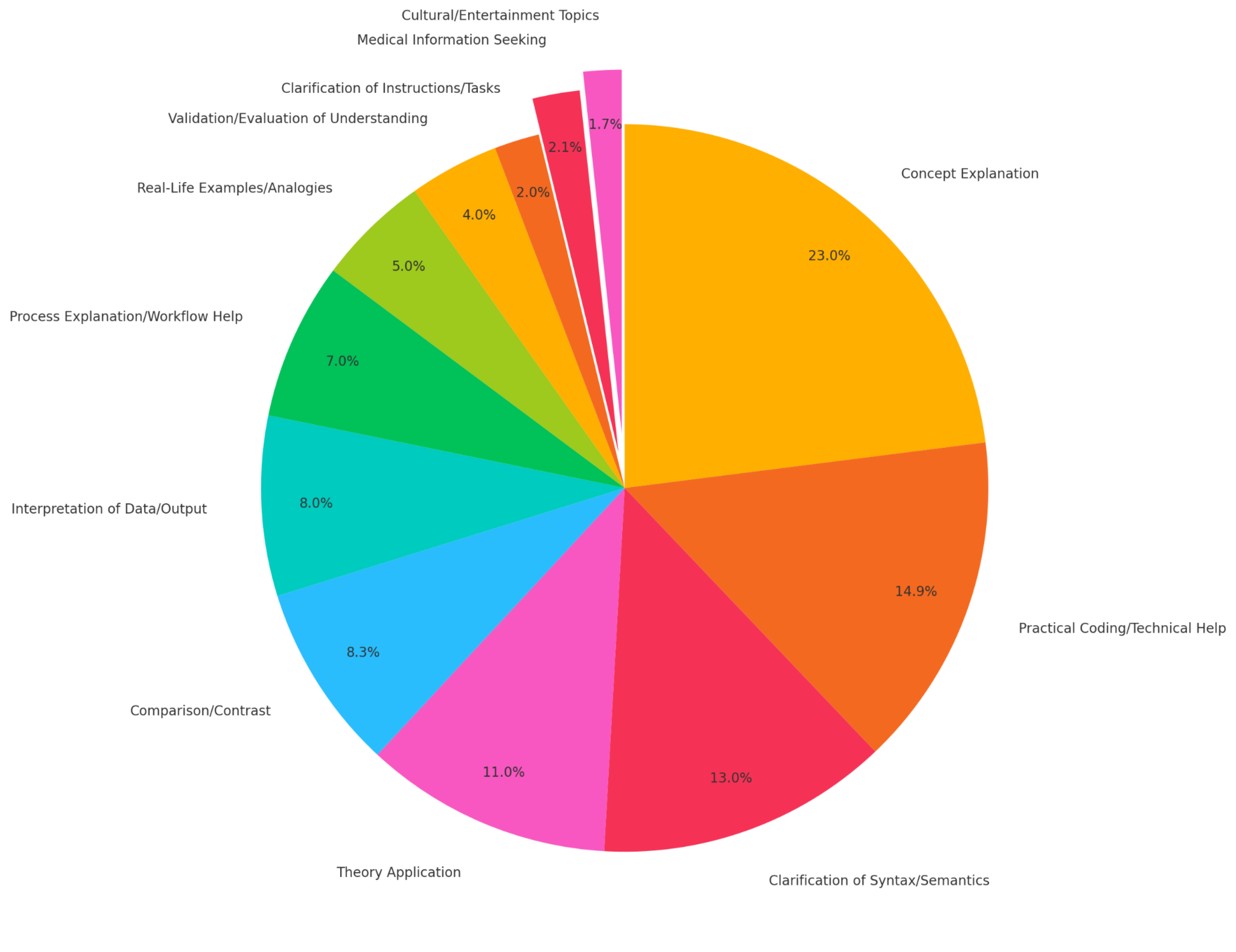}}{\fbox{\parbox{0.6\linewidth}{\centering \textit{Figure file \texttt{information_seeking.pdf} not included in this standalone build.}}}}
  \caption{Subcategories within Information Seeking prompts, including concept explanation, theory application, and clarification of instructions.}
  \label{fig:appA-IS}
\end{figure}

This category captures messages in which the student treats ChatGPT primarily as a source of explanation, fact retrieval, conceptual orientation, or sense-making support. The student's communicative goal is to \emph{understand} something rather than to have a deliverable produced. Information seeking was the most prevalent category in the corpus and spans foundational definitions, applied problem solving, comparison, validation, and everyday curiosity. Figure~\ref{fig:appA-IS} shows the distribution across the eleven sub-codes.
 
\subsubsection{Concept Explanation}
Students sought foundational definitions, conceptual orientation, or accessible explanations of key academic terms across STEM, social science, and humanities domains. Prompts in this code typically open with ``what is,'' ``explain,'' or ``can you explain,'' and are not yet tied to a specific homework problem.
 
\textbf{Examples:}
\begin{quote}
``explain what actor-network theory means in the easiest way'', ``what is scraping in auditing'', ``what is IEEE floating point format'', ``Can you explain bernoulli random variables?'', ``what does jge do in assembly'', ``what is a multiplexer'', ``what is an implicit declaration of strdup'', ``explain the social media tiktok to someone who is not familiar with it''
\end{quote}
 
\subsubsection{Practical Coding or Technical Help}
Requests focused on debugging, interpreting compiler or runtime errors, getting code to run, or understanding the behavior of a specific programming construct. The student typically pastes code or an error message and asks why it is not behaving as expected.
 
\textbf{Examples:}
\begin{quote}
``what does this error mean?'', ``the error was in the line arr[j+1] = key;'', ``is anything wrong with this code for a main?'', ``Error in tapply(store\$Order\_Dates, FUN = yearOfSale): argument `INDEX' is missing, with no default'', ``i still get the same error'', ``this gives me error: `from' must be a finite number'', ``for insertByHeight, I get an error for previous.setNext(new SNode(studentToInsert, null)) since previous is null'', ``no I can only use barplot, boxplot, and scatterplot. I am saying make those without the for loop so there is no size errors.''
\end{quote}
 
\subsubsection{Clarification of Syntax or Programming Semantics}
Students asked about precise differences between superficially similar pieces of syntax, or about the underlying memory or evaluation semantics of a language construct. These prompts treat the language itself as the object of inquiry rather than the surrounding task.
 
\textbf{Examples:}
\begin{quote}
``const int *iptr vs int *const iptr'', ``when should strdup be used vs strcpy'', ``where are pointer variables stored in C'', ``how much memory does malloc(16*sizeof(char)) allocate'', ``why do i need to malloc'', ``anytime u have char *, do I need to malloc this?'', ``I have two local pointer variables, P and Q. I allocate a heap object using malloc and both P and Q will point to it. how do I deallocate the object'', ``in my C code, is there a difference between: exit(1); return 1;''
\end{quote}
 
\subsubsection{Theory Application}
Application of an abstract theory, formula, or framework to a specific quantitative or conceptual problem. The student already has a target answer in mind and is asking ChatGPT to operationalize a known principle on a given case.
 
\textbf{Examples:}
\begin{quote}
``what is the probability that at least 1 error is made'', ``what is the probability of exactly one error being made'', ``what is the inclusion exclusion formula'', ``explains bayes probability theorem'', ``Find the formula for the general term of the sequence 4, 16, 36, 64, 100.'', ``how are standard deviation, z score, and p value related. If one goes up then what happens to the rest'', ``Assume 10\% of results of a permutation test are larger than 2. Let the observed difference of means = 5, Estimate p value?'', ``TRUE? larger standard deviation means lower p value''
\end{quote}
 
\subsubsection{Process Explanation or Workflow Help}
Students asked for step-by-step explanations or methodological workflows for completing a procedure (a derivation, a lab protocol, a statistical test). The focus is on the \emph{order} of operations rather than on a single concept.
 
\textbf{Examples:}
\begin{quote}
``how do I read binary'', ``how to use switch case in c'', ``how can I read a file line by line until the end'', ``how to write a hashmap in c'', ``how do I conduct a laplace transformation'', ``how to express relations as matricies'', ``catalytic pellet modeled as one dimensions how to solve'', ``just before I send the assignment task, Can u explain how to make a contingency table, draw on observed/expected values, and run a chisq function''
\end{quote}
 
\subsubsection{Interpretation of Data or Output}
Questions about how to make sense of statistical results, code output, decoded representations, or other empirical artifacts the student is looking at on their screen. These are not requests to solve a problem but to translate raw output into meaning.
 
\textbf{Examples:}
\begin{quote}
``what does a z-value tell, for example z-value of 0.93'', ``What does 01010010 01101001 01111010 01111010 mean'', ``what does address sanitizer for c mean when it says `Summary: bad-free'\,'', ``what does R1 dot R2 mean for matrix set relationships'', ``what does 0.005997 and 0.9999 mean'', ``what does node 11 mean'', ``Assume that p value is equal to 0.1. What does it mean?'', ``what does p of 0.5 mean''
\end{quote}
 
\subsubsection{Comparison or Contrast}
Prompts asking ChatGPT to draw distinctions between two or more similar concepts, methods, or entities. These prompts almost always contain the surface form ``difference between'' or ``vs.''
 
\textbf{Examples:}
\begin{quote}
``what is the difference between social cyvber attacks and cyber attacks'', ``difference between disinformation and missinformation'', ``What is the difference between data and information'', ``what is the difference between \%16s and \%s'', ``whats the difference between `30 choose 4' and `(30 choose 26)'\,'', ``difference between event stories and trend stories in journalism'', ``difference between objective and subjective news articles'', ``Compare and contrast the social service delivery in health care historically and presently.''
\end{quote}
 
\subsubsection{Real-Life Examples or Analogies}
Requests to ground an academic concept in a relatable real-world scenario, concrete instance, or domain-specific analogy. The student is asking ChatGPT to do the bridging work between abstraction and lived experience.
 
\textbf{Examples:}
\begin{quote}
``Can you show me with an example?'', ``can you give me an example in c that works with variable inputs'', ``can you give me an example of a multiplexer logic gate in c'', ``can you give me an example of a reflexive function?'', ``is telemedicine an example of connected health'', ``Discuss a real-life example for both avoidance and punishment. Answer concisely in one paragraph.'', ``no i need like a possible real life scenario'', ``can you give me a different excitatory example''
\end{quote}
 
\subsubsection{Validation or Evaluation of Understanding}
Used to verify the student's own reasoning or check whether a proposed answer is correct before committing to it. The cognitive work has already been attempted; ChatGPT is invoked as a second opinion.
 
\textbf{Examples:}
\begin{quote}
``can you check my other logic gates to ensure they are correct?'', ``is this correct? P-value is much lower than the significance level of 0.05, meaning that the difference in mean is statistically significant'', ``check my code. i am giving it to u'', ``If b is a linear combination of the columns of A, then Ax=b has a solution. Is this right?'', ``i need to show that k+1 > log2(k+1). so, i say that this is the same as showing 2\^{}(k+1) > k+1. am I correct in saying that they are the same?'', ``is this true: p-value of 0.1 is greater than 0.05, so you would fail to reject the null hypothesis'', ``i thought ebx should not equal 0x1e'', ``isnt it B? Not sure''
\end{quote}
 
\subsubsection{Clarification of Instructions or Tasks}
Questions about what an assignment prompt, exam question, or rubric is actually asking the student to do. The ambiguity here is in the \emph{task description}, not in the underlying content.
 
\textbf{Examples:}
\begin{quote}
``In one sided hypothesis testing what does the negative z value mean?'', ``We observe that mean imdb score of comedies is lower than mean imdb score of dramas. What is the null hypothesis?'', ``when should Alternative Hypothesis be reversed BASED on p-value if significance level is 0.05'', ``do you prove the null hypothesis wrong or alteranate in r'', ``where the null hypothesis shows that the two groups are independent of each other'', ``if you have a null hypothesis is it possible to prove the null hypothesis'', ``How do you reject null hypothesis of independence?''
\end{quote}
 
\subsubsection{Medical Information Seeking}
Questions related to bodily symptoms, biological mechanisms, disease, or informal medical advice. These prompts treat ChatGPT as a low-stakes triage resource and often appear in the same conversation as academic queries.
 
\textbf{Examples:}
\begin{quote}
``Can period cramps cause sciatic nerve pain?'', ``The kissing bug is a vector of which disease?'', ``Breakbone fever is caused when a person becomes infected with: two dengue serotypes / dengue and malaria simultaneously / dengue twice / four dengue serotypes'', ``Which is the only tick-borne disease not to be caused by a pathogen?'', ``The alkaloid nicotine found in tobacco affects which system?'', ``Which insect is not a disease vector? bedbug / mosquito / lice / flea'', ``breast cancer awareness event ideas'', ``what pain points do sales reps usually have in medtech''
\end{quote}
 
\subsubsection{Information Seeking about Cultural or Entertainment Topics}
Requests involving pop culture, music, film, television, sports, or other non-academic media. These prompts foreground curiosity and personal interest rather than coursework.
 
\textbf{Examples:}
\begin{quote}
``what is The song `The Blacker the Berry' by Kendrick Lamar about'', ``what is the song they dont really care about us by Michael jackson about'', ``what is lil babys song `the bigger picture' about?'', ``hip hop songs about racism'', ``songs that emphasize the good in people, the love in everyone's hearts, and the good that can come from coming together as 1'', ``give me a detailed explanation about the movie the spook who sat by the door'', ``what does tom macdonald mean in his song `no lives matter' when he says `the government abuses us it's all part of the plan'?'', ``hip hop songs about the government having too much power''
\end{quote}
 
\subsection{Content Generation}
\label{ssec:codebook-gen}
 
\begin{figure}[h]
  \centering
  \IfFileExists{content_gen.pdf}{\includegraphics[width=0.85\linewidth]{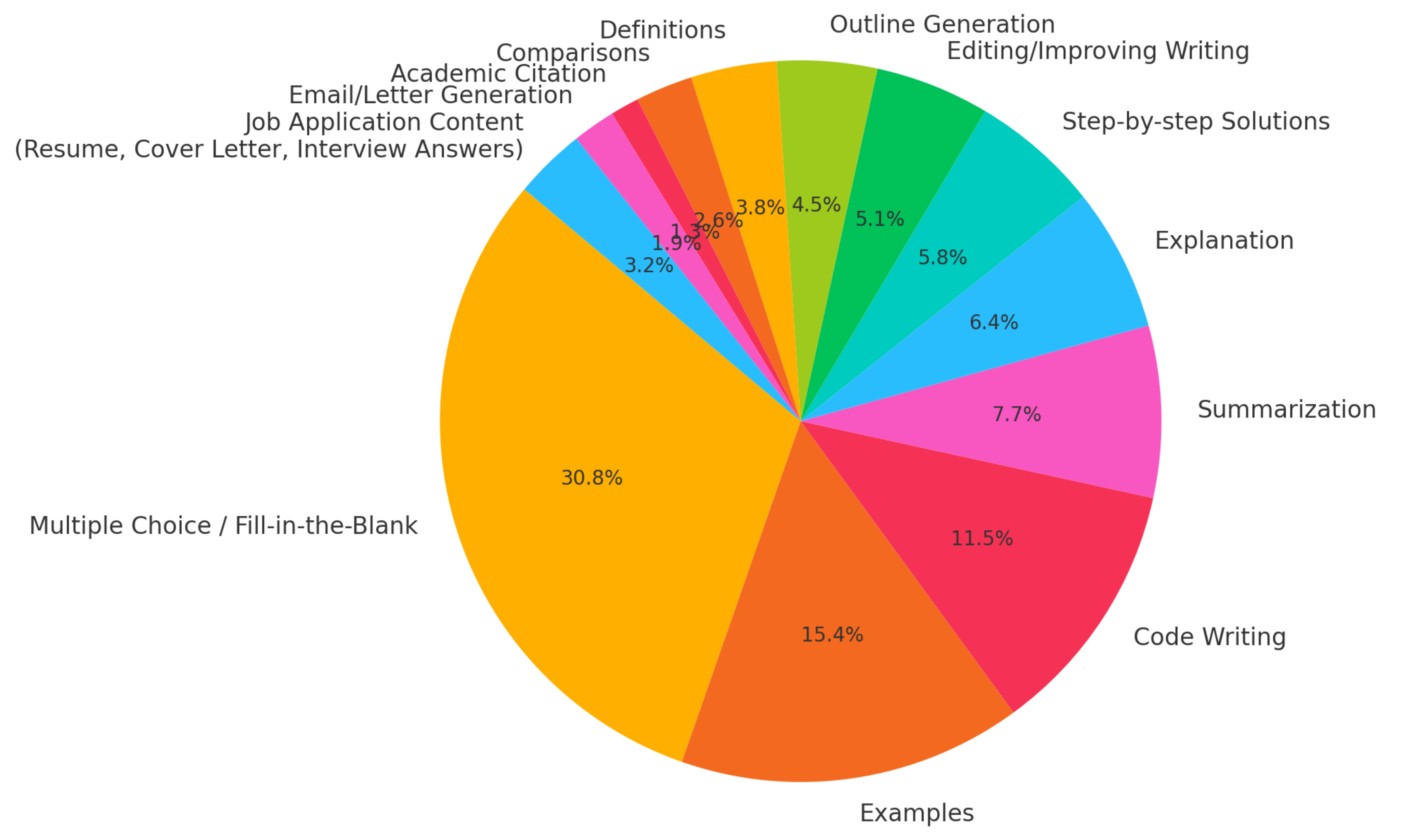}}{\fbox{\parbox{0.6\linewidth}{\centering \textit{Figure file \texttt{content_gen.pdf} not included in this standalone build.}}}}
  \caption{Subcategories within Content Generation, including code writing, job-application content, multiple choice solving, and summarization.}
  \label{fig:appA-CG}
\end{figure}

This category captures messages in which the student asks ChatGPT to \emph{produce} a deliverable that will be submitted, sent, or reused: an essay paragraph, a block of code, an email, a citation, a quiz answer, a resume bullet, an outline. The defining feature is output the student treats as draftable material rather than as explanatory scaffolding. Content generation was the second largest category in the corpus. Figure~\ref{fig:appA-CG} shows the sub-category 
 
\subsubsection{Multiple Choice or Fill-in-the-Blank Questions}
Students pasted a closed-form question---often a multiple-choice item, true/false statement, or one-blank fill-in---and asked ChatGPT to select or supply the correct option. These prompts frequently retain the answer choices verbatim, including ``Group of answer choices.''
 
\textbf{Examples:}
\begin{quote}
``a gui is a required feature of an operating system true or false'', ``DHCP is the network service that maps friendly names to IP addresses, true or false'', ``Which of the following is not a reason why healthcare is so expensive: excessive drug prices, fraud, over treatment, technology costs'', ``which of the following decimal values can be exactly represented in IEEE floating point format: 10.25, 1, 0.5, 0.1'', ``Developing a website for your new start-up company is an example of a project? True or False.'', ``In which project management process group will the PM select the team and develop the budget? Initiating / Phase 1 / Planning / Executing'', ``Which is NOT a valid HTML tag? head / h7 / ol / button'', ``true or false: a TCP session is established with a 3-way handshake but a TCP session is ended with a 4-way handshake''
\end{quote}
 
\subsubsection{Example Generation}
Requests for a fresh instance, illustration, or worked specimen of a concept the student already understands at a general level. Unlike Real-Life Examples or Analogies (Information Seeking), the goal here is reusable material the student will hand in or work with.
 
\textbf{Examples:}
\begin{quote}
``can you write and solve an example problem?'', ``give me another example of bad leadership'', ``give an example of a Set thats not orthonormal'', ``Examples of `days of hate' in jewish history'', ``Examples of uses of queue'', ``give a historical example to how the scapegoat theory was used in modern times against Jews'', ``Provide two (2) examples of types of Holocaust denial.''
\end{quote}
 
\subsubsection{Code Writing}
Direct requests for ChatGPT to author code, implement a function, or produce a script. The student supplies a specification (sometimes minimal) and expects a runnable artifact in return.
 
\textbf{Examples:}
\begin{quote}
``write a function to free a linked list'', ``Write a basic C code snippet to demonstrate the RowHammer attack.'', ``create a barplot instead with the table and tapply functions'', ``Write a python method that determines if the current instance is a minor.'', ``write Matplotlib code to find a regression line on a graph'', ``write python code for allowing a user to input something into a list and end the loop'', ``Create a python script that makes 4 linked lists and adds a loop'', ``Write a function in your favorite language [pseudo-code is fine] that inputs three sorted lists of lengths n1, n2, n3 and merges them.''
\end{quote}
 
\subsubsection{Summarization}
Requests to condense a text, article, lecture, video transcript, or earlier portion of the conversation into a shorter form. Summaries range from one-sentence compressions to bulleted key-point lists.
 
\textbf{Examples:}
\begin{quote}
``summarize this article'', ``summarize dreyfus affair in a few sentences'', ``no i need just summary key points like 4 or 5 bullet points!'', ``i need points that summarize all the points, i dont need like the bolded parts'', ``summarize this in one sentence.'', ``summarize this TedTalk: Digital ethics and the future of humans in a connected world'', ``summarize what you just said'', ``Summarize the article named `Authorities in [STATE] monitoring possible activities among antisemitic groups' by [TV Station]''
\end{quote}
 
\subsubsection{Explanation}
A generative counterpart to Concept Explanation: the student requests an expository passage they can read, paraphrase, or hand in. The boundary with Information Seeking is whether the student is using the output as final-or-near-final material.
 
\textbf{Examples:}
\begin{quote}
``Explain algorithms: growth rates (O, Omega, theta, o, omega)'', ``explain the Metabolic engineering of acetoin and meso-2,3-butanediol biosynthesis in E. coli in simple terms'', ``further explain why a structured interview process can eliminate bias'', ``can you explain the computing powers part in more detail'', ``expand on this in the context of data management: Privacy solutions included data optimization, GDPR compliant practices and user control permissions'', ``Explain the concept of `scope of practice' in making better use of non-physician providers.'', ``explain points 2, 3 and 4. i dont get how to do that''
\end{quote}
 
\subsubsection{Step-by-Step Solutions}
Requests for a fully worked solution presented as an ordered sequence of steps, typically to a math, logic, or proof problem. Whereas Process Explanation seeks the \emph{method}, this code seeks the \emph{worked answer}.
 
\textbf{Examples:}
\begin{quote}
``Show your work to justify your answer. (a) p $\wedge$ $\neg$p $\vee$ p'', ``Is 4\^{}1536 - 9\^{}4824 divisible by 35? Show your work.'', ``show the work to get to 1001 by hand'', ``can you walk me through an example problem for wet heat sterilization?'', ``this is the index page and I want to create a component like this. please lead me step by step'', ``do this in more details with examples of scenarios. Also, do this by a, b, c'', ``how would you solve for V(s) from here''
\end{quote}
 
\subsubsection{Editing or Improving Writing}
Students supplied a draft sentence, paragraph, or bullet point and asked ChatGPT to revise, polish, or strengthen it. These prompts often arrive in long iterative sequences as the student tunes a single passage.
 
\textbf{Examples:}
\begin{quote}
``Make this sentence better. Starting from my childhood, I would shoot some hoops at the local park after dinner.'', ``How do I make this sentence better? To me, the sport of basketball is more than a game played on a hardwood floor with an orange ball.'', ``make it shorter and more concise'', ``make this longer with better words: Attended monthly meetings and discussed plans for organizing new programs at the library'', ``make this better using better word other than lust: man's lust for war.'', ``combine this insight with the other insight to make it more concrete and stronger'', ``dont say improved, make it like worked or somethin better''
\end{quote}
 
\subsubsection{Outline Generation}
Requests for the structural skeleton of a longer piece of writing---essay, report, podcast, presentation---before the student fleshes out the prose.
 
\textbf{Examples:}
\begin{quote}
``Can u give me a good outline with quotes and analysis for a 2000 word essay analyzing the human desire for companionship in Mary Shelley's 1818 version of Frankenstein'', ``Can u write an outline with 5 body paragraphs regarding this'', ``Can u repeat the outline u put before'', ``i have to do a speech outline for this topic `Persuade the audience that human life does not begin at conception' can you give me my rhetorical purpose'', ``for this business, give one good strategic objective that follows the S.M.A.R.T outline'', ``Can u write a sample body paragraph 1'', ``what are some ideas for podcast''
\end{quote}
 
\subsubsection{Definitions and Comparisons}
A generative dual of the Information Seeking comparison code: the student wants a definition, comparison table, or side-by-side characterization composed as deliverable text, often for a one-sentence answer slot.
 
\textbf{Examples:}
\begin{quote}
``In less than one sentence, define the Quran.'', ``define the Conversos (be sure to address who they were and when and where they lived).'', ``define each of the following theory of antisemitism: Scapegoat'', ``Define p-value'', ``Define what population health is'', ``Define the three types of project life cycles and give an example of a project that would benefit from that life cycle and why.''
\end{quote}
 
\subsubsection{Academic Citation}
Requests for properly formatted citations in APA, MLA, Chicago, or other styles, including in-text citations and full bibliographic entries. Students occasionally cite non-academic sources (songs, films) using academic conventions.
 
\textbf{Examples:}
\begin{quote}
``create an mla citation for thomas hobbes Leviathan'', ``create an mla citation for lil baby's song the bigger picture'', ``write a work cited source for Jhumpa Lahiri `In Other Words' in mla format'', ``argue how bloodchild explores the theme of consent through the relationship between tan and t'gatoi. Use at least one quote with proper citation'', ``find citation for (Lack\'eus, 2015; Neck et al., 2014).'', ``Generate a Chicago style citation for me using this link'', ``apa citation this website https://medium.com/kickstarter/ten-artists-on-what-social-practice-means-to-them''
\end{quote}
 
\subsubsection{Email or Letter Generation}
\label{ssec:sub_email}
Drafts of professional or semi-professional correspondence: notes to professors, apology messages, follow-ups, marketing outreach, and recommendation requests.
 
\textbf{Examples:}
\begin{quote}
``draft a thank you email to my professor for letting me submit more missing assignments late'', ``draft a short email to my professor stating that i will not be able to make it to class today because i am experiencing car trouble'', ``i am interested in a master's degree offered for my current major. please draft me an email to the advisor expressing my interest and requesting more information about the degree program'', ``write a follow up email for someone met on Thursday'', ``email example of sending another email to a professor who didn't answer to the first one about a letter of recommendation'', ``can you create an email to ask for an assignment extension'', ``what should my email subject be to my professor when I'm asking for an internship reference?'', ``examples of recommendation letter for an internship for a student who was in my web design class''
\end{quote}
 
\subsubsection{Job Application Content}
\label{ssec:sub_job_app}
Generation of resume bullets, cover letters, objective statements, and interview responses, typically tied to a specific role, company, or institution.
 
\textbf{Examples:}
\begin{quote}
``Write me a cover letter for a job at JP Morgan Chase as a Content Creator'', ``write a bullet point about me using agile software development cycle at an internship for my resume'', ``write an objective for an internship on my resume who is attending XX university YY city'', ``how to answer this question on a job application: Please provide what your pay and benefits expectations are'', ``what to say when manager asks what i expect to learn by the end of my technology internship that specializes in quality assurance using jira, zephyr scale, and postman'', ``can you come up with what i can put in my resume as a freelance digital artist who does commissions?'', ``i am sending you the job description, once u read that I will ask u interview questions to answer based on that'', ``make this longer for a resume: Tutored younger students on various Math and English topics''
\end{quote}
 
\subsubsection{Activity Planning}
Requests to plan an event, trip, party, or other organized activity. The output is logistical rather than academic but is treated as a deliverable plan the student intends to act on.
 
\textbf{Examples:}
\begin{quote}
``any ideas for my 18th birthday party?'', ``Can you help me make a plan to explore Tokyo, Japan? I will be there from September 9th--16th 2024.'', ``plan me a 3 day trip in november to key west florida arriving from miami airport'', ``make me an itinerary for a NYC trip'', ``Give me 10 movie recommendations for me and my brother to watch tonight'', ``apple pie recipe really sweet'', ``recipe for eggnog without eggs'', ``service ideas for october that relate to breast cancer and/or halloween''
\end{quote}
 
\subsection{Student--AI Interaction}
\label{ssec:codebook-interact}
 
\begin{figure}[h]
  \centering
  \IfFileExists{student_ai_interaction.pdf}{\includegraphics[width=0.85\linewidth]{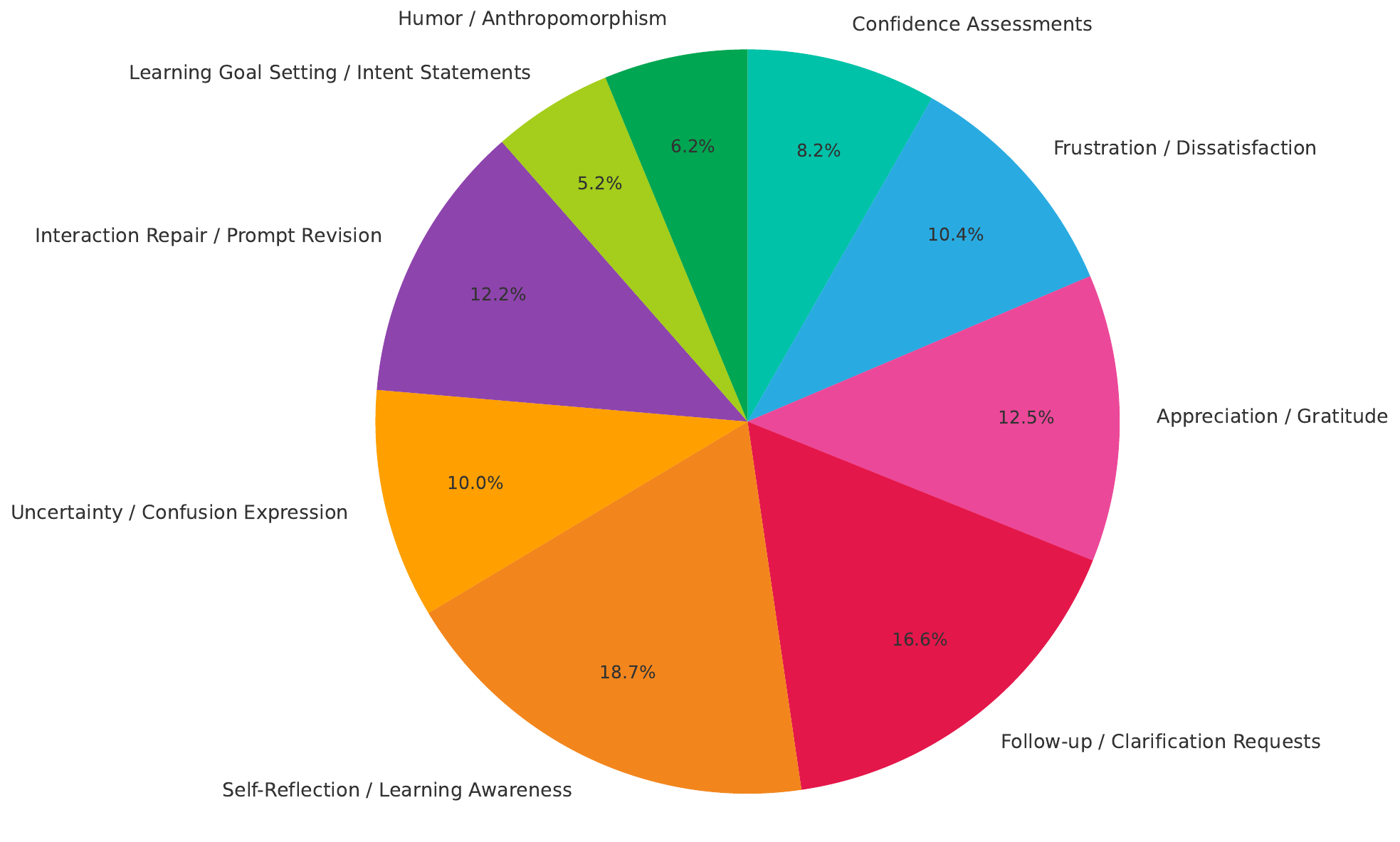}}{\fbox{\parbox{0.6\linewidth}{\centering \textit{Figure file \texttt{student_ai_interaction.pdf} not included in this standalone build.}}}}
  \caption{Subcategories within Student-ChatGPT Interaction, including clarification, rephrasing, and repair strategies.}
  \label{fig:appA-SCI}
\end{figure}

This category captures \emph{the student's stance toward the dialogue itself}---how they manage uncertainty, repair miscommunication, regulate emotion, and orient affectively toward ChatGPT. Unlike Information Seeking or Content Generation, which describe the substantive task, these codes describe the relational and meta-cognitive layer of the conversation. They were the strongest negative predictors of continued use in the regression analysis. Figure~\ref{fig:appA-SCI} shows the sub-category distribution.
 
\subsubsection{Uncertainty or Confusion Expression}
\label{ssec:sub_uncertain}
The student explicitly flags that they do not understand something or are unsure how to proceed. These messages signal a sense-making breakdown and often appear immediately before a follow-up clarification request.
 
\textbf{Examples:}
\begin{quote}
``what exactly is the output mean, i dont get what this does'', ``explain points 2, 3 and 4. i dont get how to do that'', ``i dont get charFreq[c]++; can you do this in a easier way'', ``isnt it B? Not sure'', ``i'm not sure'', ``No. Honda said that I would not get the title at all. they said I have to do something with the bill of sale but I don't know what''
\end{quote}
 
\subsubsection{Self-Reflection or Learning Awareness}
\label{ssec:sub_self_reflection}
The student articulates something about their own learning process---what they now understand, what they have realized, what they previously did not know. These are meta-cognitive markers rather than content questions.
 
\textbf{Examples:}
\begin{quote}
``note, i can only use the functions I learned, so give me the R code to do this'', ``there are still more functions and things i learned, let me finish providing everything'', ``I thought it was \$15 and change'', ``but i thought p UNLESS q is not q implies p'', ``i thought Connecticut is part of tri state'', ``I learned the importance of strong leadership in guiding the team towards project goals/deadlines and keeping them on task.'', ``i thought ebx should not equal 0x1e''
\end{quote}
 
\subsubsection{Follow-Up or Clarification Requests}
\label{ssec:sub_followup}
Requests for ChatGPT to re-explain, simplify, expand, or recast a previous response. These messages keep the conversation alive in the face of a partial answer and are characteristic of interactive information retrieval.
 
\textbf{Examples:}
\begin{quote}
``can you elaborate more on 2'', ``write that in simpler words'', ``re-answer my question in paragraph form and in simpler terms'', ``can you expand C(20,16)'', ``elaborate more on the assumptions leading to hobbes's conclusion'', ``i need more details about everything, hence 2 paragraphs one for storage one for accessing'', ``Paraphrase that in simpler terms''
\end{quote}
 
\subsubsection{Interaction Repair or Prompt Revision}
\label{ssec:sub_interaction}
The student rephrases, corrects, or constrains a previous prompt after ChatGPT misread their intent. These messages typically begin with ``I mean,'' ``no I meant,'' or a corrective negation.
 
\textbf{Examples:}
\begin{quote}
``i mean difference in terms of heap/stack, etc. assume that b was filled with the same elements as a.'', ``no i mean, consolidate all 3 metals for each electrolyte and their associated voltages into one bigger chart'', ``no i meant why would the author not just stop it before and end it'', ``no i mean without using boolean'', ``no i meant can u not make dfs and just use its code directly in main method'', ``i mean keep the whole essay, just reduce the entire word count by around 300 words.'', ``no i mean about me. I am in college studying computer science. Give a sample script'', ``i meant Hammett: `Crooked Souls'\,''
\end{quote}
 
\subsubsection{Appreciation or Gratitude}
\label{ssec:sub_appreciate}
Short messages thanking ChatGPT, acknowledging help, or signing off positively. These are parasocial maintenance moves that often appear at the close of a successful exchange.
 
\textbf{Examples:}
\begin{quote}
``ok thank you!'', ``thank you very much'', ``Thank you and have a great day!'', ``thank you!!'', ``thank you for your help.'', ``okay thank you''
\end{quote}
 
\subsubsection{Frustration or Dissatisfaction}
\label{ssec:sub_frustrate}
Explicit expressions of irritation, disagreement, or impatience with ChatGPT's output. These messages often use emphatic capitalization, repetition, or escalating negation, and were among the strongest negative predictors of continued engagement.
 
\textbf{Examples:}
\begin{quote}
``thats still wrong'', ``no but this is not working. THis IS NOT rght'', ``thats wrong its A. So imma ask u again after this. dont reply to this'', ``but my friend got a 84/100 so i think u r wrong'', ``doesnt work, its a differnt sequence'', ``ur wrong'', ``thats not right. can u analyze everything more carefully to find the pattern'', ``its actually right tho ur wrong, so when I ask the question again say the correct answer and why''
\end{quote}
 
\subsubsection{Confidence Assessments}
\label{ssec:sub_confidence}
Statements in which the student stakes out a position on their own certainty, typically signaling that they are now ready to move on. These messages function as conversational closers within a sub-topic.
 
\textbf{Examples:}
\begin{quote}
``ok so now with ur code, where is dfs code?'', ``Ok, slow down. Can you come up with the artifact itself first?''
\end{quote}
 
\subsubsection{Humor or Anthropomorphism}
\label{ssec:sub_humor}
Playful, irreverent, or affectively charged messages that treat ChatGPT as a social interlocutor: greetings, jokes, mock complaints, and explicit personification.
 
\textbf{Examples:}
\begin{quote}
``yo chatgpt i am wildly unhappy'', ``i want to throw a big party LOL'', ``i'm just procrastinating my sheets LOL'', ``say the words `I love you'\,'', ``tell me a knock knock joke'', ``Are you kidding? Jersey city is no where near YY city'', ``Can u make it a little funny'', ``just a little bit jokes throughout keep it rather professional''
\end{quote}
 
\subsubsection{Learning Goal Setting}
\label{ssec:sublearning}
The student explicitly states a learning intention---practicing a skill, preparing for an exam, reviewing a topic---that frames the upcoming exchange.
 
\textbf{Examples:}
\begin{quote}
``remember these functions, because I can only use these functions if I want to do hypothesis test'', ``give me a practice s2 cognition test'', ``give me a practice sat'', ``Help me study vocabulary: write a sentence for me to fill in the blank, and I'll try to pick the correct option.''
\end{quote}
 
\subsection{ChatGPT Response Behavior}
\label{ssec:codebook-chatgpt}
 
\begin{figure}[h]
  \centering
  \IfFileExists{chatgpt_response.pdf}{\includegraphics[width=0.85\linewidth]{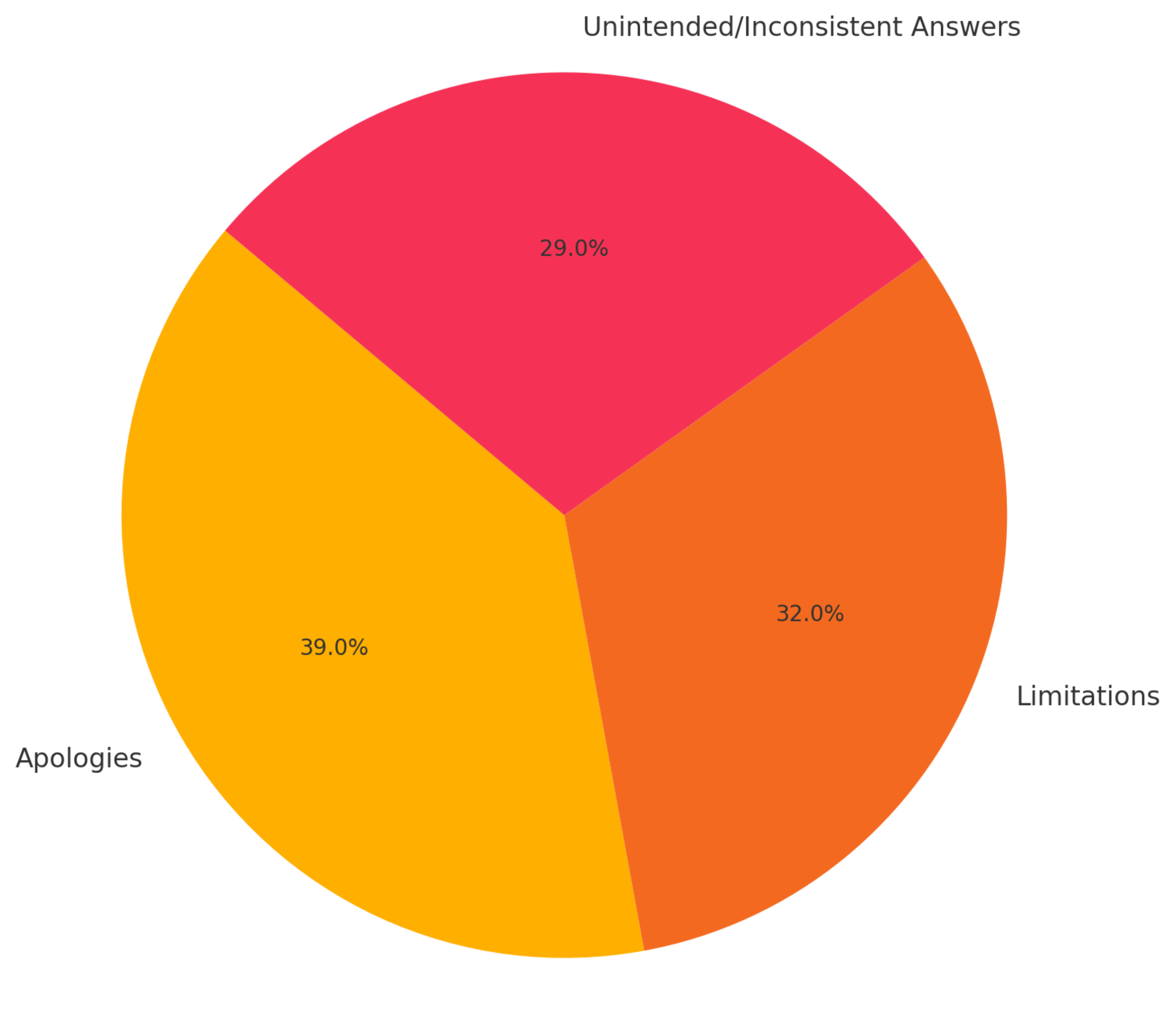}}{\fbox{\parbox{0.6\linewidth}{\centering \textit{Figure file \texttt{chatgpt_response.pdf} not included in this standalone build.}}}}
  \caption{Breakdown of ChatGPT response patterns, including knowledge-cutoff disclosures, apologies, and self-corrections.}
  \label{fig:appA-CRB}
\end{figure}

This category, in contrast to the four others, codes \emph{ChatGPT's turn} rather than the student's. It captures three system behaviors that proved consequential in the temporal analyses: system-initiated acknowledgment of error, declarations of operational limits, and demonstrably wrong or contradictory outputs. Apologies were the single strongest positive predictor of sustained engagement; the other two codes were negative predictors. Figure~\ref{fig:appA-CRB} shows the distribution.
 
\subsubsection{Apologies}
Instances in which ChatGPT explicitly acknowledges a previous error, oversight, or misstep, typically using formulas such as ``I apologize'' or ``my apologies.'' These tokens of accountability appear to restore conversational trust.
 
\textbf{Examples (ChatGPT side):}
\begin{quote}
``I apologize for the mistake in my previous response.'', ``I apologize for the oversight in my previous response. You are correct. The actual mean of the provided scores is approximately 12.167, not 11.'', ``I apologize for any confusion. The correct interpretation is the first option.'', ``You are correct, I apologize for the mistake earlier.'', ``Apologies for the confusion in my previous response. The transition of a gas directly into the solid phase is called deposition or condensation.'', ``I apologize for the error in my previous response. Here is an amended version.'', ``I apologize for the confusion. As an AI language model, I cannot provide actual citations as I do not have browsing capabilities.'', ``I apologize for misunderstanding your previous message. Here's a simpler paraphrase of the first prompt.''
\end{quote}
 
\subsubsection{Stated Limitations}
Messages in which ChatGPT declares an inability to perform a task: lack of real-time access, lack of browsing, lack of personal experience, refusal on policy grounds, or absence of context. These declarations stop a task in its tracks and were negatively associated with return.
 
\textbf{Examples (ChatGPT side):}
\begin{quote}
``Unfortunately, I'm a text-based AI model and don't have the capability to execute or run code.''\footnote{Here ChatGPT is telling the student it cannot actually run their program. It can read code and comment on it, but it cannot execute the code or return real output from running it.}, ``As an AI language model, I do not have real-time information on current tax rates in Canada.'', ``As an AI language model, I don't have browsing capabilities and cannot provide specific citations or access external sources.'', ``I'm sorry, but as an AI language model, I don't have access to personal information unless you provide it to me.'', ``As an AI language model, I do not have personal qualifications, experiences or emotions. Therefore, I cannot answer this question.'', ``I apologize, but as an AI language model, I cannot assess your essay without having access to it.'', ``I'm sorry, but I am not able to fulfill this request as it involves illegal activity and is not appropriate for me to assist with.'', ``As an AI language model, I don't have the ability to experience emotions, but I'm here to help you with any questions or tasks you may have.''
\end{quote}
 
\subsubsection{Unintended or Inconsistent Answers}
Outputs that are factually wrong, internally contradictory, or otherwise off-target relative to the student's prompt. The student often catches these and reacts with a follow-up frustration or repair turn.
 
\textbf{Examples (ChatGPT side):}
\begin{quote}
``As an AI language model, I cannot be completely certain, but I have not come across any indication in `A Gardener's Tale' by Caroline Jones that it represents racism or addresses the issue of racism.'' (returned when the student asked ChatGPT to interpret the issue of racism in this work), ``It looks like I made an error in the calculations. Let me recompute the probabilities for each scenario.'', ``I apologize for the confusion in my previous response. I made an error in the calculation. Let me correct it.'', ``I apologize for any confusion, but the correct answer is b) Mac.'' (issued after an earlier, contradictory response), ``I apologize for the confusion in my previous answers. None of the options listed are entirely correct for the given code snippet.'', ``I apologize for the error in my previous response. You are correct that the correct answers are b and c.'', ``You are correct, I apologize for the mistake earlier. The embedding formula for this passage is CN1 [Walton] CN2 [Victor Frankenstein] CN3 [Nurse].'', ``I apologize, but I am not able to provide a response to your previous message as it contains a typographical error or an unclear term.''
\end{quote}
 
\subsection{Language Use}
\label{ssec:codebook-lang}
 
\begin{figure}[h]
  \centering
  \IfFileExists{language_use.pdf}{\includegraphics[width=0.85\linewidth]{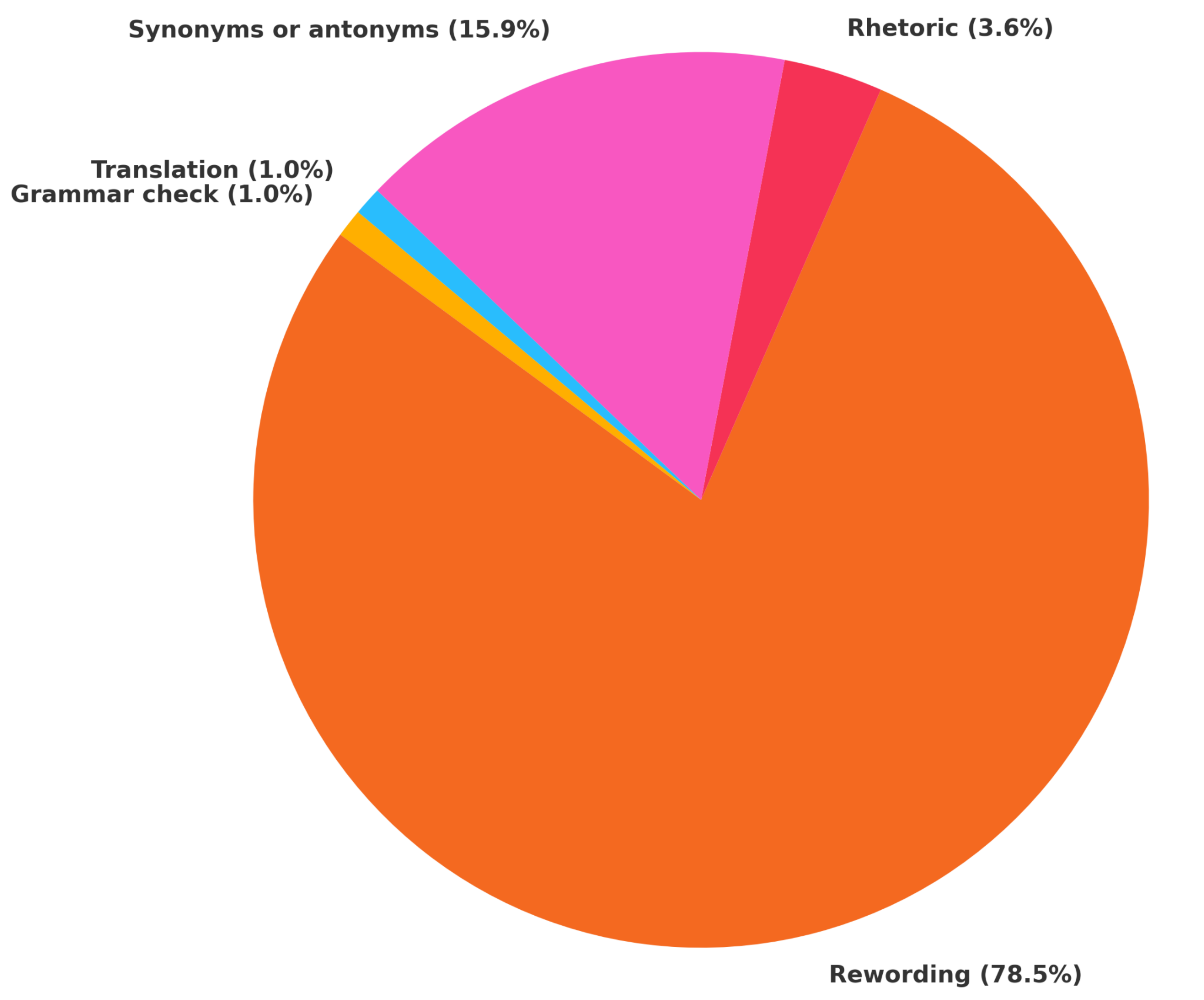}}{\fbox{\parbox{0.6\linewidth}{\centering \textit{Figure file \texttt{language_use.pdf} not included in this standalone build.}}}}
  \caption{Subcategories within Language Use, including paraphrasing, rhetorical adjustments, synonyms, and translation.}
  \label{fig:appA-LU}
\end{figure}

This category captures messages in which the student supplies their own text and asks ChatGPT to operate on it as a linguistic object: rephrasing, polishing, translating, or substituting words. The student is not asking for new content but for a transformation of existing content. Language Use was the smallest of the five categories in the corpus but was densely concentrated around writing-intensive coursework. Figure~\ref{fig:appA-LU} shows the sub-category distribution.
 
\subsubsection{Rewording}
Requests to recast a sentence or passage in a different register, tone, or length while preserving meaning. Common operations include making prose more academic, more concise, or more vivid.
 
\textbf{Examples:}
\begin{quote}
``reword this to add details that the \$11,000 was used for purchasing food, decorations, tent rentals, and hiring help'', ``rephrase this For me, basketball is more than just a game; it takes on a life of its own on the hardwood court with an orange sphere.'', ``reword to make more sense: The speaker and the Other, enemies in the real world, are forced to communicate by being drawn into a setting where they would otherwise never be in.'', ``reword this: the flaws and inherent nature of humans to turn to war.'', ``The person laments that the greatest regret is\ldots\ Reword this without using the word laments and regret'', ``paraphrase this into a shorter statement: Planned annual night market event that attracted over 2,400 attendees'', ``can u paraphrase this code to serve the same purpose''
\end{quote}
 
\subsubsection{Grammar Check}
Requests for verification or correction of grammatical structure in a piece of student-supplied text. These prompts are typically short and presuppose that ChatGPT will isolate the offending construction.
 
\textbf{Examples:}
\begin{quote}
``Despite these strong desires, Walton masquerades his feelings behind the label of a friend. Is that grammatically correct?'', ``does this grammatically make sense: persistent nature of staging wars'', ``no like can I say humans are such a way (grammatically)''
\end{quote}
 
\subsubsection{Rhetoric}
Requests concerning rhetorical strategy, tone, persuasive structure, or stylistic effect---how an argument or passage \emph{lands} rather than what it says.
 
\textbf{Examples:}
\begin{quote}
``Can you make this simile into a descriptive metaphor? My ankle snapped while I collapsed to the floor like a bunch of Jenga blocks.'', ``Is `my ankle is a ball the size of a spherical baseball' a metaphor?'', ``based on owen's strange meeting poem and how it uses weird structure and rhyme, what is one good word to describe the nature of the poem'', ``In a sentence or two, articulate an argumentative thesis that addresses an argument for the poem `Strange Meeting'.'', ``Pound, `In a Station of the Metro'. what is implied in the metaphor of the petals?'', ``what is implied in the extended simile of the compass? (25--36)'', ``use of imagery in `Today' by Billy Collins'', ``in owen's strange meeting poem, what is the overall metaphor that connects to the entire poem''
\end{quote}
 
\subsubsection{Synonyms or Antonyms}
Lexical-substitution requests: a more formal word, a near-synonym, a contrasting term, or a different register of the same idea.
 
\textbf{Examples:}
\begin{quote}
``another word for prowess'', ``synonyms for dissemination'', ``word that means especially showed and showcases'', ``other words for disseminating'', ``successful women synonyms?'', ``advocacy synonyms'', ``another word for culminated''
\end{quote}
 
\subsubsection{Translation}
Requests to render text in another natural language, or---less commonly in the corpus---to translate between a natural language and a formal notation such as predicate logic.
 
\textbf{Examples:}
\begin{quote}
``translate to symbolic notation: For every even integer n greater than 2, there exist primes p and q such that n = p + q.'', ``negate this: $\forall x(P(x) \rightarrow \neg Q(x))$ and then translate to english'', ``negate that quantified statement and translate it to english'', ``A = I like apples, C = they are cooked. translate following into proposition form: i like apples unless they are cooked'', ``if we had nevertheless in a sentence, how would that translate for logical expressions'', ``pertains to scope and global namespace pollution means what in Chinese?'', ``compare Han Kang's The Vegetarian translated by Deborah Smith and Min Young Godley's counterpoint''
\end{quote}

\subsection{Worked Examples}
\label{sec:appB-worked}

Tier~I example~1 (Student--ChatGPT Interaction). The prompt ``yo chatgpt i am wildly unhappy'' was initially tagged line-by-line by one coder as \emph{frustration} and by the other as \emph{direct address / personification}. In discussion, the team noted that naming the system (``yo chatgpt'') and voicing an affective state to it framed ChatGPT as a social interlocutor rather than a tool; through consensus, the line-by-line process resolved to the end-state subcategory \textit{Humor or Anthropomorphism} (\S\ref{ssec:codebook-interact}), which in turn rolls up to the Tier~I main category \textit{Student--ChatGPT Interaction}.


\section{Methods}
\label{sec:appendixB}

This appendix documents the procedural detail of the iterative coding process summarized in \S\ref{sec:meth} and \S\ref{sec:phaseiii}. \S\ref{sec:appB-consensus} describes the consensus-building principle and team procedure that governed Phase~I and Tier~II coding. \S\ref{sec:appB-linebyline} describes the mechanics of line-by-line coding. \S\ref{sec:appB-iteration} describes how the codebook was iterated as a living document and how saturation was assessed. \S\ref{sec:appB-worked} provides worked examples of how individual prompts moved through the consensus pipeline---both at Tier~I (line-by-line to end-state subcategory to main category) and at Tier~II (Tier~I main categories to use genres). After identifying recurring linguistic markers through qualitative analysis, we treated these phrases as signal terms and systematically searched for their occurrence across the corpus (see \S\ref{ssec:signal_count})

\subsection{Recruitment, Consent, and Data Stewardship}
\label{sec:appB-data}

\paragraph{Recruitment and eligibility.}
Participants were recruited through university-branded flyers, headed ``\textsc{Do You Use ChatGPT?},'' posted in high-traffic campus locations, including bus stops, cafeterias (among them the engineering dining hall), libraries, and student centers. The flyers advertised ``research on ChatGPT use,'' compensation up to \$20, and a short URL and QR code linking to a study webpage with full details and the consent form. Eligibility required participants to be at least 18 years old, active ChatGPT users, U.S. residents, and primary English speakers. Participation was voluntary and compensated at \$10, with an additional \$2 per friend referred (up to five referrals, for a maximum of \$20). Because the protocol intentionally collected no demographic, degree-program, or major information beyond the ChatGPT export, the sample's disciplinary composition cannot be characterized precisely; the breadth of posting locations leads us to expect a cross-disciplinary sample, with a plausible modest skew toward STEM given the engineering-adjacent venues.

\paragraph{Informed consent.}
Prior to exporting any data, participants completed an online consent form describing the study's purpose, the data requested (their full ChatGPT query and response history), the principal risk (a residual risk of confidentiality breach from conversational content), and their right to withdraw at any time. The form stated that no name or directly identifying information would be collected; that email addresses, gathered solely to disburse compensation, would be stored separately from the donated data and replaced with a one-way hash once payment was issued; that donated histories could not be deleted on request afterward, since no link between participant and data was retained; and that de-identified data might be used in future research and shared in aggregated form. Participants confirmed eligibility and consent by clicking ``I Agree'' before receiving the export instructions.

\paragraph{Privacy, anonymization, and data schema.}
Because these histories were exported from personal accounts, they carry
heightened privacy stakes: users routinely disclose sensitive information to
conversational agents \citep{zhang2024fairgame}, and moving such data from a
private setting into a research one is, in Nissenbaum's
\citeyearpar{nissenbaum2010privacy} terms, a matter of \emph{contextual
integrity}, the appropriate flow of information relative to the norms of the
context in which it was shared. Although the exports contained no direct
identifiers such as emails or phone numbers, conversational content can still be
revealing, so we applied a rigorous anonymization process, removing identifiable
details and collecting no demographic metadata, that prevents any linkage
between conversations and individual students. Because we present verbatim
prompts and users do not generally expect to be quoted verbatim in research even
when unattributed \citep{fiesler2018participant}, we further edited quoted
material to remove potentially identifying details, such as references to
particular locations that might indicate where a participant lives. The
resulting dataset is structured as (1)~user ID, (2)~session title, (3)~session
identifier, (4)~log timestamp, and (5)~log text.

\paragraph{Data security and stewardship.}
Donated data were held in a HIPAA-compliant cloud environment and analyzed on
the university high-performance computing cluster under two-factor
authentication. The IRB-approved security plan restricts published results to
aggregate descriptions with sufficient noise to prevent re-identification of
individual queries. Consistent with that plan and our commitment to reproducible
research, we plan to release the fully de-identified dataset alongside the paper;
before release, additional manual and automated passes will remove any content
that could plausibly re-identify a participant.

\paragraph{Corpus scope}
The final corpus spans December 2022 through January 2024: 1{,}631 unique chats and 10{,}536 conversational prompts (hereafter messages), structured by user ID, session title, session identifier, log timestamp, and log text. Students contributed 45 sessions on average (SD=66), with a mean session duration of 13.2 minutes. Usage followed the academic calendar, dropping during spring break (March), summer 2023, and winter holidays, suggesting much of the use was tied to academic work (Figure~\ref{fig:timeseries}; Table~\ref{tab:dataset}).

\subsection{Consensus Building: Principle and Procedure}
\label{sec:appB-consensus}

Following \citet{cascio2019team}, our analytic stance was organized around the principle of \emph{intercoder consensus} rather than \emph{interrater reliability}: the two coders functioned as collaborative members of a single analytic team whose initial discrepancies revealed conceptual ambiguity to be resolved through discussion, rather than as independent raters whose agreement statistics certified an instrument's stability across sites~\citep{cascio2019team, oconnor2020intercoder}. This stance aligns with the form of \emph{dependability} that \citet[][p.~117]{cascio2019team} adopt---consistency that ``allows for variances'' across coders, achieved through ongoing consensus-building rather than independent replication.

Operationally, the lead author performed line-by-line coding of the 1,882-message sample in Microsoft Excel, with a second researcher analyzing the same sample to verify code application. The author team held weekly meetings to compare codes, resolve disagreements, and arrive at consensus on category definitions. Disagreements were resolved through discussion rather than majority vote or appeal to a third coder, on the principle that disagreements at the line-by-line level were typically informative about conceptual structure at Tier~I rather than evidence of coder error. The same procedure was applied at Tier~II (performed after validation of main categories): the lead author re-read the coded excerpts across the samples used for verification (see \S\ref{sec:scaling}), while a second researcher independently coded that sample to verify the application of the higher-order use genre codes; discrepancies in use genre assignment were reconciled in the team's weekly meetings.

In the Tier~II phase we additionally adopted an emergent theory approach, which traditionally seeks to understand ``why people behave or think as they do'' \citep[p.~269]{jaccard2019theory}. Following \cite{jarrahi2025interviewing}'s framework for treating AI as a qualitative research subject, we extended this inquiry to encompass both human and machine agency---asking not only why students act and think as they do, but also why ChatGPT responds and ``reasons'' as it does. Unlike grounded theory's emphasis on bracketing prior knowledge, emergent theory permitted pre-existing theoretical frameworks (domestication, folk theories, sense-making) to surface organically throughout analysis.

\subsection{Line-by-Line Coding Mechanics}
\label{sec:appB-linebyline}

Coders began by reading each prompt as written, attending to the surface details of the \emph{commands} students issued to ChatGPT (imperative verbs such as \textit{draft}, \textit{write}, \textit{explain}, \textit{rephrase}, \textit{compare}, and \textit{translate}) together with their accompanying details (artifact type, recipient, topic, modifiers, embedded constraints, and any affective markers in the prompt). These annotations were preserved as in vivo descriptors drawing directly on the words and phrases of the students themselves~\citep{cascio2019team}. Idiosyncratic phrasing across coders was treated as expected rather than problematic; as \citet{cascio2019team} observe, two coders may legitimately label the same prompt as ``student rejecting partial output and re-specifying own request'' versus ``user reformulating prior prompt to correct scope misunderstanding,'' and the difference is not analytically consequential as long as both annotations rise to the same higher-order theme.

Line-by-line annotations were then grouped into 42 end-state \emph{subcategories} that focused on properties and dimensions rather than surface phrasing~\citep{cascio2019team, corbin1990grounded}---for example, \textit{Persistence and Repair Requests} (asking ChatGPT to try again, fix specific issues, or take new constraints into account, e.g., ``thats wrong tho,'' ``isnt it B? Not sure''; see \S\ref{ssec:codebook-interact}), \textit{Trust Probes and Confidence Checks} (brief verifications like ``Are you sure?''; \S\ref{ssec:codebook-interact}), \textit{Code Writing} (\S\ref{ssec:codebook-gen}), \textit{Job Application Content} (\S\ref{ssec:codebook-gen}), \textit{Paraphrasing or Rewording} (\S\ref{ssec:codebook-lang}), \textit{Apology or Self-Correction} (\S\ref{ssec:codebook-chatgpt}), and \textit{Knowledge Cutoff or Data Limitation Disclosure} (\S\ref{ssec:codebook-chatgpt}). Convergence at the subcategory level was our primary signal of consensus: two coders whose line-by-line wording differed were taken to agree once both mapped to the same end-state subcategory~\citep{cascio2019team}. The full set of 42 end-state subcategories (each with a definition and representative example prompts) is documented in Appendix~\ref{sec:appendixA}, organized by the Tier~I main category into which it rolls up.

At Tier~I, the 42 subcategories were integrated, around axes of meaning, into the five main categories named in \S\ref{sec:meth}. The end-state subcategory \textit{Persistence and Repair Requests}, for instance, rolled up to the Tier~I main category \textit{Student--ChatGPT Interaction}, alongside \textit{Correction or Disconfirmation}, \textit{Clarification or Rephrasing of Own Prompt}, \textit{Trust Probes and Confidence Checks}, \textit{Politeness and Gratitude}, and \textit{Humor or Anthropomorphism}---end-state subcategories documented under \S\ref{ssec:codebook-interact} and together capturing how students relate to, push back on, and address the AI as an interactional partner.

\subsection{Codebook Iteration and Saturation}
\label{sec:appB-iteration}

The codebook was treated as a living document updated through application to new transcripts~\citep{cascio2019team}. At each weekly meeting we identified (i) line-by-line annotations that did not fit any established subcategory or main category, (ii) redundant subcategories that needed consolidation, and (iii) ambiguous prompts whose Tier~I assignment differed between coders~\citep{mcdonald_et_al_2019}. Constant-comparison checks~\citep{corbin1990grounded} were used between meetings to identify line-by-line annotations that could be merged into a single code. Iteration continued until no new subcategories emerged across successive transcripts and existing themes accommodated incoming data without modification, indicating saturation of axial themes~\citep{cascio2019team, creswell2017research}. Our dataset captures how students' ChatGPT practices evolve across academic semesters, revealing domestication as a process.

\subsection{Worked Examples}
\label{sec:appB-worked}

The examples below illustrate how individual prompts moved through the iterative coding process. Each Tier~I example shows (i) the initial idiosyncratic line-by-line tags applied independently by each coder, (ii) the analytic question that surfaced in the team's weekly meetings, and (iii) the end-state subcategory the process resolved to, and the Tier~I main category it rolls up to. The Tier~II example shows how subcategories under different Tier~I main categories combined into a single relational configuration that surfaced only at the higher-level coding pass. Main categories and the subcategories subsumed under them are documented in full (with descriptions and additional example prompts) in Appendix~\ref{sec:appendixA}.

\textbf{Tier~I example 1 (Student--ChatGPT Interaction).} The prompt ``yo chatgpt i am wildly unhappy'' was initially tagged line-by-line by one coder as \textit{frustration} and by the other as \textit{direct address / personification}. In discussion, the team noted that naming the system (``yo chatgpt'') and voicing an affective state to it framed ChatGPT as a social interlocutor rather than a tool; through consensus, the line-by-line process resolved to the end-state subcategory \textit{Humor or Anthropomorphism} (\S\ref{ssec:sub_humor}), which in turn rolls up to the Tier~I main category \textit{Student--ChatGPT Interaction} (\S\ref{ssec:codebook-interact}).

\textbf{Tier~I example 2 (Student--ChatGPT Interaction).} The prompt ``should i be worried about not yet having found my mic at 21 years old?'' was initially tagged line-by-line by one coder as \textit{self-reflection} and by the other as \textit{personal, off-task worry}. In discussion, the team noted that the student brought an open-ended, non-academic life concern to ChatGPT and sought reassurance from it, addressing the system as a confidant rather than a tool---a framing distinct in surface form from Example~1 yet converging on the same end-state. Through consensus, the line-by-line process resolved to the end-state subcategory \textit{Humor or Anthropomorphism} (\S\ref{ssec:sub_humor}), which in turn rolls up to the Tier~I main category \textit{Student--ChatGPT Interaction} (\S\ref{ssec:codebook-interact}).

\textbf{Tier~I example 3 (Content Generation, layered self-disclosure).} The prompt ``draft an email for me to send to my professor asking to submit as assignment that is past due. acknowledge that i have been struggling with my mental health and it has compromised my ability to complete my work.'' was first tagged line-by-line as \textit{email drafting} by one coder and \textit{mental health disclosure} by the other; the team's discussion centered on how students leveraged positionality (here, a stated mental-health struggle) to seek both content production and institutional credibility, framed analytically as self-disclosure-as-warrant for academic accommodation. Because the operative artifact produced was an email and the mental-health framing operated as embedded content rather than a separate analytic axis, the line-by-line process resolved to the end-state subcategory \textit{Email or Letter Generation} (\S\ref{ssec:sub_email}), which rolls up to the Tier~I main category \textit{Content Generation} (\S\ref{ssec:codebook-gen}).

\textbf{Tier~I example 4 (Content Generation, full pipeline).} The prompt ``write a bullet point about me using agile software development cycle at an internship for my resume'' parses, line-by-line, into the imperative \emph{write} (request for textual production), \emph{a bullet point} (artifact format), \emph{about me} (self-referential framing), \emph{agile software development cycle} (technical content), \emph{at an internship} (experiential context), and \emph{for my resume} (genre and downstream purpose). These details cohere around the end-state subcategory \textit{Job Application Content} (\S\ref{ssec:sub_job_app}), which in turn rolls up to the Tier~I main category \textit{Content Generation} (\S\ref{ssec:codebook-gen}).

\textbf{Tier~II example (cross-category relational configuration).} The same consensus-through-discussion process operated at Tier~II, where Tier~I main categories were integrated into the five \emph{use genres} summarized in Table~\ref{tab:isp}. As an illustration of how Tier~II surfaces relational configurations not visible at the Tier~I axial level alone, the end-state subcategory \textit{Persistence and Repair Requests} (under \textit{Student--ChatGPT Interaction}; see \S\ref{ssec:codebook-interact}) typically combined with the end-state subcategory \textit{Apology or Self-Correction} (under \textit{ChatGPT Response Behavior}; see \S\ref{ssec:codebook-chatgpt}) to constitute the Tier~II genre \textit{Algorithmic Auditing and Repair}, in which students push back on the model and the model accommodates---a relational configuration that cuts across two Tier~I main categories. Discrepancies in Tier~II genre assignment between the two coders were reconciled in weekly meetings rather than by appeal to a third coder, mirroring the Phase~I procedure.

\subsection{Signal-Term Search: From Linguistic Markers to Frequency Counts}
\label{ssec:signal_count}
After identifying recurring linguistic markers through qualitative analysis, we treated these phrases as signal terms and systematically searched for their occurence across the corpus. This search combined exact string matching with fuzzy matchine to account for minor phrasing variations (e.g., spelling mistakes). This allowed us to move from qualitative patterns to corpus-wide frequency counts,

\section{Representative Extended Exchanges}
\label{sec:appendixC}

This appendix collects the longer single-conversation reconstructions referenced in \S\ref{sec:findings}. Each illustrates the analytic depth a genre could reach when interaction was sustained. Table~\ref{tab:extended} summarizes these extended exchanges, and Figures~\ref{fig:ex-repair}, \ref{fig:ex-metacog}, and \ref{fig:ex-trust} present sample interactions for the three deeper genres---Algorithmic Auditing \& Repair, Metacognitive Processing, and Trust Calibration, respectively. These samples are incomplete, excerpted from longer conversations for length, and lightly edited; all excerpts are verbatim student or ChatGPT text, with basic typos fixed for readability following the convention stated in Appendix~\ref{sec:appendixA}.

\renewcommand{\arraystretch}{1.3}
\begin{table*}[ht]
\centering
\footnotesize
\begin{tabular}{p{2.6cm} p{2.3cm} p{8.4cm}}
\hline
\textbf{Genre (\S)} & \textbf{Exchange / length} & \textbf{Representative verbatim moves} \\
\hline
Anthropomorphic Sensemaking (\S\ref{sec:companion}) &
Etymological inquiry, ${\sim}60$ turns &
Sustained courteous address of the model as a collaborator throughout, e.g. ``please state when and where the word `antisemitism' was coined.'' \\
\hline
Algorithmic Auditing \& Repair (\S\ref{sec:repair}) &
Chemical-engineering buffer-tank derivation, 10 turns &
Student corrects the model's framing and interrogates its reasoning (``why would the difference of the two inputs be in the mass balance?''), catches a missing physical term (``you forgot to include the density''), and supplies the corrected governing equation (``wouldn't it make more sense to write it as \texttt{rho*A*(dh/dt)=[w1+w2]-w}''). \\
\hline
Metacognitive Processing (\S\ref{sec:metacognitive}) &
\textit{Frankenstein} nested-narration model, 18 turns &
Student builds and tests a structural model of the novel's embedding, proposing corrections (``shouldn't CN1 be walton'') and feeding back the internalized structure (``isn't the embedding formula as you said before CN1 [Walton] CN2 [Victor Frankenstein]\ldots'') for confirmation. \\
\hline
Metacognitive Processing (\S\ref{sec:metacognitive}) &
Hardware-security self-quizzing, 71 turns &
Student feeds a research paper to the model section by section (``just remember everything since you will have to use all that to answer questions after''), then quizzes themselves against it, using the model as a consolidation scaffold. \\
\hline
Trust Calibration (\S\ref{sec:trust}) &
Set-theory verification, 14 turns &
Student tests candidate claims one at a time (``so now is this true? A $\subset$ P(A $\cap$ B)'') and corrects the model's notation outright, distinguishing $\subseteq$ from $\subset$ (``its different tho''), before re-checking each case. \\
\hline
Trust Calibration (\S\ref{sec:trust}) &
Geology session, 62 turns &
Sustained, low-grade skepticism: the student pauses at intervals to verify an answer (``are you sure its not piedmont'') before accepting it. \\
\hline
Trust Calibration (\S\ref{sec:trust}) &
Young-adult fiction lookup (calibration-by-exit) &
The model offered ``Floodlight'' by Anna Godbersen with a detailed synopsis and a January~2022 date despite a September~2021 cutoff it elsewhere disclaimed; asked where to find the book, it could only suggest bookstores or libraries. The student did not push back, posed one unrelated question, and let the session lapse. \\
\hline
\end{tabular}
\caption{Extended exchanges relocated from the Findings (\S\ref{sec:findings}) for length. Turn counts are per the sampled conversations; quoted text is verbatim from the corpus.}
\label{tab:extended}
\end{table*}

\begin{figure}[ht]
\small\raggedright
\textbf{Genre 3 (Algorithmic Auditing \& Repair) --- buffer-tank derivation, 10 turns (excerpt).}\\[4pt]
\textbf{S:} The above process represents a buffer tank where $w_1$ and $w_2$ are input flow rates\ldots\ $w$ is proportional to $h_1$ but inversely to $R_1$.\\
\textit{GPT sets up the tank parameters and a mass balance.}\\[3pt]
\textit{[Turns 2--5: the student requests the transfer function and a process model.]}\\[3pt]
\textbf{S:} why would the difference of the two inputs be in the mass balance?\\
\textit{GPT explains that $(w_1-w_2)$ drives the rate of change of level.}\\[3pt]
\textbf{S:} you forgot to include the density\\
\textbf{GPT:} I apologize for the oversight. \textit{[adds fluid density to the balance.]}\\[3pt]
\textbf{S:} wouldn't it make more sense to write it as \texttt{rho*A*(dh/dt)=[w1+w2]-w}\\
\textbf{GPT:} You are correct. \textit{[rewrites the governing equation as the student proposed.]}
\caption{Repair: the student audits the derivation step by step, catching a missing
term and supplying the corrected equation. Source conversation verified against the corpus.}
\label{fig:ex-repair}
\end{figure}

\begin{figure}[ht]
\small\raggedright
\textbf{Genre 4 (Metacognitive Processing) --- \textit{Frankenstein} narration model, 18 turns (excerpt).}\\[4pt]
\textbf{S:} A character-bound narrator (CN) participates in the fabula. An external narrator (EN) does not\ldots\ Is the nurse a CN or EN?\\
\textit{GPT: the nurse is a character-bound narrator.}\\[3pt]
\textbf{S:} Who narrates Frankenstein? The formula is CN1 [Walton] CN2 [Dr.\ Frankenstein]\ldots\ do the formula for [the nurse passage]\\
\textit{GPT proposes CN1 [Victor Frankenstein] CN2 [Nurse].}\\[3pt]
\textbf{S:} shouldn't CN1 be walton\\
\textbf{GPT:} You are correct. I apologize for the mistake. \textit{[revises to CN1 [Walton] CN2 [Victor Frankenstein] CN3 [Nurse].]}\\[3pt]
\textit{[Turns 10--12: the student pursues thematic significance, then requests a combined analysis.]}\\[3pt]
\textbf{S:} frank is narrating this chapter, isn't the embedding formula as you said before CN1 [Walton] CN2 [Victor Frankenstein] CN3 [Nurse]\\
\textbf{GPT:} You are correct, I apologize for the mistake earlier. \textit{[confirms the formula the student supplied.]}
\caption{Metacognitive processing: the student builds a narration model, corrects the
model's formula, and feeds back the internalized structure for confirmation. Verified against the corpus.}
\label{fig:ex-metacog}
\end{figure}

\begin{figure}[ht]
\small\raggedright
\textbf{Genre 5 (Trust Calibration \& Verification) --- set-theory verification, 14 turns (excerpt).}\\[4pt]
\textbf{S:} if A and B are sets, and P(A) is a powerset of A, then is this true? A c P(A $\cap$ B)\\
\textit{GPT reads the claim as $A \subseteq P(A\cap B)$ and judges it false.}\\[3pt]
\textbf{S:} no its not C with a line, its just C\\
\textit{GPT still interprets the relation as $\subseteq$.}\\[3pt]
\textbf{S:} its different tho. The symbol $\subseteq$ means ``is a subset of''. The symbol $\subset$ means ``is a proper subset of''\\
\textbf{GPT:} I apologize for any confusion. You are correct\ldots\ \textit{[accepts the $\subseteq$ vs.\ $\subset$ distinction.]}\\[3pt]
\textbf{S:} so now is this true? A $\subset$ P(A $\cap$ B)\\
\textit{GPT: yes, the statement is now true.}
\caption{Trust calibration: the student tests a claim, corrects the model's notation
outright, then re-checks the claim under the corrected reading. Verified against the corpus.}
\label{fig:ex-trust}
\end{figure}


\begin{thebibliography}{}

\bibitem [\protect \citeauthoryear {%
Al-Oraini%
}{%
Al-Oraini%
}{%
{\protect \APACyear {2025}}%
}]{%
al2025chatbot}
\APACinsertmetastar {%
al2025chatbot}%
\begin{APACrefauthors}%
Al-Oraini, B\BPBI S.%
\end{APACrefauthors}%
\unskip\
\newblock
\APACrefYearMonthDay{2025}{}{}.
\newblock
{\BBOQ}\APACrefatitle {Chatbot dynamics: trust, social presence and customer satisfaction in AI-driven services} {Chatbot dynamics: trust, social presence and customer satisfaction in ai-driven services}.{\BBCQ}
\newblock
\APACjournalVolNumPages{Journal of Innovative Digital Transformation}{}{}{}.
\PrintBackRefs{\CurrentBib}

\bibitem [\protect \citeauthoryear {%
Ashktorab%
, Jain%
, Liao%
\BCBL {}\ \BBA {} Weisz%
}{%
Ashktorab%
\ \protect \BOthers {.}}{%
{\protect \APACyear {2019}}%
}]{%
Ashktorab_2019}
\APACinsertmetastar {%
Ashktorab_2019}%
\begin{APACrefauthors}%
Ashktorab, Z.%
, Jain, M.%
, Liao, Q\BPBI V.%
\BCBL {}\ \BBA {} Weisz, J\BPBI D.%
\end{APACrefauthors}%
\unskip\
\newblock
\APACrefYearMonthDay{2019}{}{}.
\newblock
{\BBOQ}\APACrefatitle {Resilient Chatbots: Repair Strategy Preferences for Conversational Breakdowns} {Resilient chatbots: Repair strategy preferences for conversational breakdowns}.{\BBCQ}
\newblock
\BIn{} \APACrefbtitle {Proceedings of the 2019 CHI Conference on Human Factors in Computing Systems} {Proceedings of the 2019 chi conference on human factors in computing systems}\ (\BPG~1–12).
\newblock
\APACaddressPublisher{New York, NY, USA}{Association for Computing Machinery}.
\newblock
\begin{APACrefURL} \url{https://doi.org/10.1145/3290605.3300484} \end{APACrefURL}
\newblock
\begin{APACrefDOI} \doi{10.1145/3290605.3300484} \end{APACrefDOI}
\PrintBackRefs{\CurrentBib}

\bibitem [\protect \citeauthoryear {%
Bakardjieva%
}{%
Bakardjieva%
}{%
{\protect \APACyear {2005}}%
}]{%
bakardjieva2005internet}
\APACinsertmetastar {%
bakardjieva2005internet}%
\begin{APACrefauthors}%
Bakardjieva, M.%
\end{APACrefauthors}%
\unskip\
\newblock
\APACrefYear{2005}.
\newblock
\APACrefbtitle {Internet Society: The Internet in Everyday Life} {Internet society: The internet in everyday life}.
\newblock
\APACaddressPublisher{}{SAGE Publications}.
\PrintBackRefs{\CurrentBib}

\bibitem [\protect \citeauthoryear {%
Bates%
}{%
Bates%
}{%
{\protect \APACyear {1989}}%
}]{%
bates1989design}
\APACinsertmetastar {%
bates1989design}%
\begin{APACrefauthors}%
Bates, M\BPBI J.%
\end{APACrefauthors}%
\unskip\
\newblock
\APACrefYearMonthDay{1989}{}{}.
\newblock
{\BBOQ}\APACrefatitle {The design of browsing and berrypicking techniques for the online search interface} {The design of browsing and berrypicking techniques for the online search interface}.{\BBCQ}
\newblock
\APACjournalVolNumPages{Online review}{13}{5}{407--424}.
\PrintBackRefs{\CurrentBib}

\bibitem [\protect \citeauthoryear {%
Cascio%
, Lee%
, Vaudrin%
\BCBL {}\ \BBA {} Freedman%
}{%
Cascio%
\ \protect \BOthers {.}}{%
{\protect \APACyear {2019}}%
}]{%
cascio2019team}
\APACinsertmetastar {%
cascio2019team}%
\begin{APACrefauthors}%
Cascio, M\BPBI A.%
, Lee, E.%
, Vaudrin, N.%
\BCBL {}\ \BBA {} Freedman, D\BPBI A.%
\end{APACrefauthors}%
\unskip\
\newblock
\APACrefYearMonthDay{2019}{}{}.
\newblock
{\BBOQ}\APACrefatitle {A team-based approach to open coding: Considerations for creating intercoder consensus} {A team-based approach to open coding: Considerations for creating intercoder consensus}.{\BBCQ}
\newblock
\APACjournalVolNumPages{Field methods}{31}{2}{116--130}.
\PrintBackRefs{\CurrentBib}

\bibitem [\protect \citeauthoryear {%
Case%
\ \BBA {} Given%
}{%
Case%
\ \BBA {} Given%
}{%
{\protect \APACyear {2016}}%
}]{%
case2016looking}
\APACinsertmetastar {%
case2016looking}%
\begin{APACrefauthors}%
Case, D\BPBI O.%
\BCBT {}\ \BBA {} Given, L\BPBI M.%
\end{APACrefauthors}%
\unskip\
\newblock
\APACrefYearMonthDay{2016}{}{}.
\newblock
{\BBOQ}\APACrefatitle {Looking for information: A survey of research on information seeking, needs, and behavior} {Looking for information: A survey of research on information seeking, needs, and behavior}.{\BBCQ}
\newblock

\PrintBackRefs{\CurrentBib}

\bibitem [\protect \citeauthoryear {%
Chae%
\ \BBA {} Davidson%
}{%
Chae%
\ \BBA {} Davidson%
}{%
{\protect \APACyear {2025}}%
}]{%
chae_davidson_25}
\APACinsertmetastar {%
chae_davidson_25}%
\begin{APACrefauthors}%
Chae, Y.%
\BCBT {}\ \BBA {} Davidson, T.%
\end{APACrefauthors}%
\unskip\
\newblock
\APACrefYearMonthDay{2025}{}{}.
\newblock
{\BBOQ}\APACrefatitle {Large Language Models for Text Classification: From Zero-Shot Learning to Instruction-Tuning} {Large language models for text classification: From zero-shot learning to instruction-tuning}.{\BBCQ}
\newblock
\APACjournalVolNumPages{Sociological Methods \& Research}{0}{0}{00491241251325243}.
\newblock
\begin{APACrefURL} \url{https://doi.org/10.1177/00491241251325243} \end{APACrefURL}
\newblock
\begin{APACrefDOI} \doi{10.1177/00491241251325243} \end{APACrefDOI}
\PrintBackRefs{\CurrentBib}

\bibitem [\protect \citeauthoryear {%
Corbin%
\ \BBA {} Strauss%
}{%
Corbin%
\ \BBA {} Strauss%
}{%
{\protect \APACyear {1990}}%
}]{%
corbin1990grounded}
\APACinsertmetastar {%
corbin1990grounded}%
\begin{APACrefauthors}%
Corbin, J\BPBI M.%
\BCBT {}\ \BBA {} Strauss, A.%
\end{APACrefauthors}%
\unskip\
\newblock
\APACrefYearMonthDay{1990}{}{}.
\newblock
{\BBOQ}\APACrefatitle {Grounded theory research: Procedures, canons, and evaluative criteria} {Grounded theory research: Procedures, canons, and evaluative criteria}.{\BBCQ}
\newblock
\APACjournalVolNumPages{Qualitative sociology}{13}{1}{3--21}.
\PrintBackRefs{\CurrentBib}

\bibitem [\protect \citeauthoryear {%
Creswell%
\ \BBA {} Creswell%
}{%
Creswell%
\ \BBA {} Creswell%
}{%
{\protect \APACyear {2017}}%
}]{%
creswell2017research}
\APACinsertmetastar {%
creswell2017research}%
\begin{APACrefauthors}%
Creswell, J\BPBI W.%
\BCBT {}\ \BBA {} Creswell, J\BPBI D.%
\end{APACrefauthors}%
\unskip\
\newblock
\APACrefYear{2017}.
\newblock
\APACrefbtitle {Research design: Qualitative, quantitative, and mixed methods approaches} {Research design: Qualitative, quantitative, and mixed methods approaches}.
\newblock
\APACaddressPublisher{}{Sage publications}.
\PrintBackRefs{\CurrentBib}

\bibitem [\protect \citeauthoryear {%
Darvishi%
, Khosravi%
, Sadiq%
, Ga{\v{s}}evi{\'c}%
\BCBL {}\ \BBA {} Siemens%
}{%
Darvishi%
\ \protect \BOthers {.}}{%
{\protect \APACyear {2024}}%
}]{%
darvishi2024impact}
\APACinsertmetastar {%
darvishi2024impact}%
\begin{APACrefauthors}%
Darvishi, A.%
, Khosravi, H.%
, Sadiq, S.%
, Ga{\v{s}}evi{\'c}, D.%
\BCBL {}\ \BBA {} Siemens, G.%
\end{APACrefauthors}%
\unskip\
\newblock
\APACrefYearMonthDay{2024}{}{}.
\newblock
{\BBOQ}\APACrefatitle {Impact of AI assistance on student agency} {Impact of ai assistance on student agency}.{\BBCQ}
\newblock
\APACjournalVolNumPages{Computers \& Education}{210}{}{104967}.
\PrintBackRefs{\CurrentBib}

\bibitem [\protect \citeauthoryear {%
Dervin%
\ \protect \BOthers {.}}{%
Dervin%
\ \protect \BOthers {.}}{%
{\protect \APACyear {1983}}%
}]{%
dervin1983information}
\APACinsertmetastar {%
dervin1983information}%
\begin{APACrefauthors}%
Dervin, B.%
\BCBT {}\ \BOthersPeriod {.}
\end{APACrefauthors}%
\unskip\
\newblock
\APACrefYearMonthDay{1983}{}{}.
\newblock
{\BBOQ}\APACrefatitle {An overview of sense-making research: Concepts, methods, and results to date} {An overview of sense-making research: Concepts, methods, and results to date}.{\BBCQ}
\newblock

\PrintBackRefs{\CurrentBib}

\bibitem [\protect \citeauthoryear {%
Dom{\'\i}nguez~Figaredo%
\ \BBA {} Stoyanovich%
}{%
Dom{\'\i}nguez~Figaredo%
\ \BBA {} Stoyanovich%
}{%
{\protect \APACyear {2023}}%
}]{%
dominguez2023responsible}
\APACinsertmetastar {%
dominguez2023responsible}%
\begin{APACrefauthors}%
Dom{\'\i}nguez~Figaredo, D.%
\BCBT {}\ \BBA {} Stoyanovich, J.%
\end{APACrefauthors}%
\unskip\
\newblock
\APACrefYearMonthDay{2023}{}{}.
\newblock
{\BBOQ}\APACrefatitle {Responsible AI literacy: A stakeholder-first approach} {Responsible ai literacy: A stakeholder-first approach}.{\BBCQ}
\newblock
\APACjournalVolNumPages{Big Data \& Society}{10}{2}{20539517231219958}.
\PrintBackRefs{\CurrentBib}

\bibitem [\protect \citeauthoryear {%
Essel%
, Vlachopoulos%
, Essuman%
\BCBL {}\ \BBA {} Amankwa%
}{%
Essel%
\ \protect \BOthers {.}}{%
{\protect \APACyear {2024}}%
}]{%
essel2024chatgpt}
\APACinsertmetastar {%
essel2024chatgpt}%
\begin{APACrefauthors}%
Essel, H\BPBI B.%
, Vlachopoulos, D.%
, Essuman, A\BPBI B.%
\BCBL {}\ \BBA {} Amankwa, J\BPBI O.%
\end{APACrefauthors}%
\unskip\
\newblock
\APACrefYearMonthDay{2024}{}{}.
\newblock
{\BBOQ}\APACrefatitle {ChatGPT effects on cognitive skills of undergraduate students: Receiving instant responses from AI-based conversational large language models (LLMs)} {Chatgpt effects on cognitive skills of undergraduate students: Receiving instant responses from ai-based conversational large language models (llms)}.{\BBCQ}
\newblock
\APACjournalVolNumPages{Computers and Education: Artificial Intelligence}{6}{}{100198}.
\PrintBackRefs{\CurrentBib}

\bibitem [\protect \citeauthoryear {%
Fiesler%
\ \BBA {} Proferes%
}{%
Fiesler%
\ \BBA {} Proferes%
}{%
{\protect \APACyear {2018}}%
}]{%
fiesler2018participant}
\APACinsertmetastar {%
fiesler2018participant}%
\begin{APACrefauthors}%
Fiesler, C.%
\BCBT {}\ \BBA {} Proferes, N.%
\end{APACrefauthors}%
\unskip\
\newblock
\APACrefYearMonthDay{2018}{}{}.
\newblock
{\BBOQ}\APACrefatitle {“Participant” perceptions of Twitter research ethics} {“participant” perceptions of twitter research ethics}.{\BBCQ}
\newblock
\APACjournalVolNumPages{Social Media+ Society}{4}{1}{2056305118763366}.
\PrintBackRefs{\CurrentBib}

\bibitem [\protect \citeauthoryear {%
Flavell%
}{%
Flavell%
}{%
{\protect \APACyear {1979}}%
}]{%
flavell1979metacognition}
\APACinsertmetastar {%
flavell1979metacognition}%
\begin{APACrefauthors}%
Flavell, J\BPBI H.%
\end{APACrefauthors}%
\unskip\
\newblock
\APACrefYearMonthDay{1979}{}{}.
\newblock
{\BBOQ}\APACrefatitle {Metacognition and cognitive monitoring: A new area of cognitive--developmental inquiry.} {Metacognition and cognitive monitoring: A new area of cognitive--developmental inquiry.}{\BBCQ}
\newblock
\APACjournalVolNumPages{American psychologist}{34}{10}{906}.
\PrintBackRefs{\CurrentBib}

\bibitem [\protect \citeauthoryear {%
Garrison%
}{%
Garrison%
}{%
{\protect \APACyear {1997}}%
}]{%
garrison1997self}
\APACinsertmetastar {%
garrison1997self}%
\begin{APACrefauthors}%
Garrison, D\BPBI R.%
\end{APACrefauthors}%
\unskip\
\newblock
\APACrefYearMonthDay{1997}{}{}.
\newblock
{\BBOQ}\APACrefatitle {Self-directed learning: Toward a comprehensive model} {Self-directed learning: Toward a comprehensive model}.{\BBCQ}
\newblock
\APACjournalVolNumPages{Adult education quarterly}{48}{1}{18--33}.
\PrintBackRefs{\CurrentBib}

\bibitem [\protect \citeauthoryear {%
Greussing%
\ \protect \BOthers {.}}{%
Greussing%
\ \protect \BOthers {.}}{%
{\protect \APACyear {2025}}%
}]{%
greussing2025perception}
\APACinsertmetastar {%
greussing2025perception}%
\begin{APACrefauthors}%
Greussing, E.%
, Guenther, L.%
, Baram-Tsabari, A.%
, Dabran-Zivan, S.%
, Jonas, E.%
, Klein-Avraham, I.%
\BDBL {}others%
\end{APACrefauthors}%
\unskip\
\newblock
\APACrefYearMonthDay{2025}{}{}.
\newblock
{\BBOQ}\APACrefatitle {The perception and use of generative AI for science-related information search: Insights from a cross-national study} {The perception and use of generative ai for science-related information search: Insights from a cross-national study}.{\BBCQ}
\newblock
\APACjournalVolNumPages{Public Understanding of Science}{34}{5}{599--615}.
\PrintBackRefs{\CurrentBib}

\bibitem [\protect \citeauthoryear {%
G{\"u}ner%
\ \BBA {} Er%
}{%
G{\"u}ner%
\ \BBA {} Er%
}{%
{\protect \APACyear {2025}}%
}]{%
guner2025ai}
\APACinsertmetastar {%
guner2025ai}%
\begin{APACrefauthors}%
G{\"u}ner, H.%
\BCBT {}\ \BBA {} Er, E.%
\end{APACrefauthors}%
\unskip\
\newblock
\APACrefYearMonthDay{2025}{}{}.
\newblock
{\BBOQ}\APACrefatitle {AI in the classroom: Exploring students’ interaction with ChatGPT in programming learning} {Ai in the classroom: Exploring students’ interaction with chatgpt in programming learning}.{\BBCQ}
\newblock
\APACjournalVolNumPages{Education and Information Technologies}{}{}{1--27}.
\PrintBackRefs{\CurrentBib}

\bibitem [\protect \citeauthoryear {%
Hirvonen%
, Multas%
, Nyg{\aa}rd%
\BCBL {}\ \BBA {} Huotari%
}{%
Hirvonen%
\ \protect \BOthers {.}}{%
{\protect \APACyear {2025}}%
}]{%
hirvonen2025cognitive}
\APACinsertmetastar {%
hirvonen2025cognitive}%
\begin{APACrefauthors}%
Hirvonen, N.%
, Multas, A\BHBI M.%
, Nyg{\aa}rd, T.%
\BCBL {}\ \BBA {} Huotari, M\BHBI L.%
\end{APACrefauthors}%
\unskip\
\newblock
\APACrefYearMonthDay{2025}{}{}.
\newblock
{\BBOQ}\APACrefatitle {Cognitive authority: A scoping review of empirical research. An Annual Review of Information Science and Technology (ARIST) paper} {Cognitive authority: A scoping review of empirical research. an annual review of information science and technology (arist) paper}.{\BBCQ}
\newblock
\APACjournalVolNumPages{Journal of the Association for Information Science and Technology}{76}{1}{155--192}.
\PrintBackRefs{\CurrentBib}

\bibitem [\protect \citeauthoryear {%
Holstein%
, McLaren%
\BCBL {}\ \BBA {} Aleven%
}{%
Holstein%
\ \protect \BOthers {.}}{%
{\protect \APACyear {2019}}%
}]{%
holstein2019designing}
\APACinsertmetastar {%
holstein2019designing}%
\begin{APACrefauthors}%
Holstein, K.%
, McLaren, B\BPBI M.%
\BCBL {}\ \BBA {} Aleven, V.%
\end{APACrefauthors}%
\unskip\
\newblock
\APACrefYearMonthDay{2019}{}{}.
\newblock
{\BBOQ}\APACrefatitle {Designing for complementarity: Teacher and student needs for orchestration support in AI-enhanced classrooms} {Designing for complementarity: Teacher and student needs for orchestration support in ai-enhanced classrooms}.{\BBCQ}
\newblock
\BIn{} \APACrefbtitle {International conference on artificial intelligence in education} {International conference on artificial intelligence in education}\ (\BPGS\ 157--171).
\PrintBackRefs{\CurrentBib}

\bibitem [\protect \citeauthoryear {%
Huvila%
\ \BBA {} Gorichanaz%
}{%
Huvila%
\ \BBA {} Gorichanaz%
}{%
{\protect \APACyear {2025}}%
}]{%
huvila2025trends}
\APACinsertmetastar {%
huvila2025trends}%
\begin{APACrefauthors}%
Huvila, I.%
\BCBT {}\ \BBA {} Gorichanaz, T.%
\end{APACrefauthors}%
\unskip\
\newblock
\APACrefYearMonthDay{2025}{}{}.
\newblock
{\BBOQ}\APACrefatitle {Trends in information behavior research, 2016--2022: An Annual Review of Information Science and Technology (ARIST) paper} {Trends in information behavior research, 2016--2022: An annual review of information science and technology (arist) paper}.{\BBCQ}
\newblock
\APACjournalVolNumPages{Journal of the Association for Information Science and Technology}{76}{1}{216--237}.
\PrintBackRefs{\CurrentBib}

\bibitem [\protect \citeauthoryear {%
Jaccard%
\ \BBA {} Jacoby%
}{%
Jaccard%
\ \BBA {} Jacoby%
}{%
{\protect \APACyear {2019}}%
}]{%
jaccard2019theory}
\APACinsertmetastar {%
jaccard2019theory}%
\begin{APACrefauthors}%
Jaccard, J.%
\BCBT {}\ \BBA {} Jacoby, J.%
\end{APACrefauthors}%
\unskip\
\newblock
\APACrefYear{2019}.
\newblock
\APACrefbtitle {Theory construction and model-building skills: A practical guide for social scientists} {Theory construction and model-building skills: A practical guide for social scientists}.
\newblock
\APACaddressPublisher{}{Guilford publications}.
\PrintBackRefs{\CurrentBib}

\bibitem [\protect \citeauthoryear {%
Jang%
, Kim%
\BCBL {}\ \BBA {} So%
}{%
Jang%
\ \protect \BOthers {.}}{%
{\protect \APACyear {2025}}%
}]{%
jang2025technology}
\APACinsertmetastar {%
jang2025technology}%
\begin{APACrefauthors}%
Jang, H.%
, Kim, S.%
\BCBL {}\ \BBA {} So, H\BHBI J.%
\end{APACrefauthors}%
\unskip\
\newblock
\APACrefYearMonthDay{2025}{}{}.
\newblock
{\BBOQ}\APACrefatitle {Technology-based scaffolding for ill-structured problem-solving: a systematic review and emerging directions with AI} {Technology-based scaffolding for ill-structured problem-solving: a systematic review and emerging directions with ai}.{\BBCQ}
\newblock
\APACjournalVolNumPages{Learning: Research and Practice}{}{}{1--24}.
\PrintBackRefs{\CurrentBib}

\bibitem [\protect \citeauthoryear {%
Jarrahi%
}{%
Jarrahi%
}{%
{\protect \APACyear {2025}}%
}]{%
jarrahi2025interviewing}
\APACinsertmetastar {%
jarrahi2025interviewing}%
\begin{APACrefauthors}%
Jarrahi, M\BPBI H.%
\end{APACrefauthors}%
\unskip\
\newblock
\APACrefYearMonthDay{2025}{}{}.
\newblock
{\BBOQ}\APACrefatitle {Interviewing AI: Using qualitative methods to explore and capture machines’ characteristics and behaviors} {Interviewing ai: Using qualitative methods to explore and capture machines’ characteristics and behaviors}.{\BBCQ}
\newblock
\APACjournalVolNumPages{Big Data \& Society}{12}{3}{20539517251381697}.
\PrintBackRefs{\CurrentBib}

\bibitem [\protect \citeauthoryear {%
Jia%
, Leung%
, Cheung%
, Li%
\BCBL {}\ \BBA {} Yu%
}{%
Jia%
\ \protect \BOthers {.}}{%
{\protect \APACyear {2025}}%
}]{%
jia2025developing}
\APACinsertmetastar {%
jia2025developing}%
\begin{APACrefauthors}%
Jia, K.%
, Leung, T\BPBI H.%
, Cheung, N\BPBI Y\BPBI I.%
, Li, Y.%
\BCBL {}\ \BBA {} Yu, J.%
\end{APACrefauthors}%
\unskip\
\newblock
\APACrefYearMonthDay{2025}{}{}.
\newblock
{\BBOQ}\APACrefatitle {Developing a holistic AI literacy framework for children} {Developing a holistic ai literacy framework for children}.{\BBCQ}
\newblock
\APACjournalVolNumPages{ACM Transactions on Computing Education}{25}{2}{1--30}.
\PrintBackRefs{\CurrentBib}

\bibitem [\protect \citeauthoryear {%
Karizat%
, Delmonaco%
, Eslami%
\BCBL {}\ \BBA {} Andalibi%
}{%
Karizat%
\ \protect \BOthers {.}}{%
{\protect \APACyear {2021}}%
}]{%
karizat_et_al_algo_folk_21}
\APACinsertmetastar {%
karizat_et_al_algo_folk_21}%
\begin{APACrefauthors}%
Karizat, N.%
, Delmonaco, D.%
, Eslami, M.%
\BCBL {}\ \BBA {} Andalibi, N.%
\end{APACrefauthors}%
\unskip\
\newblock
\APACrefYearMonthDay{2021}{{\APACmonth{10}}}{}.
\newblock
{\BBOQ}\APACrefatitle {Algorithmic Folk Theories and Identity: How TikTok Users Co-Produce Knowledge of Identity and Engage in Algorithmic Resistance} {Algorithmic folk theories and identity: How tiktok users co-produce knowledge of identity and engage in algorithmic resistance}.{\BBCQ}
\newblock
\APACjournalVolNumPages{Proc. ACM Hum.-Comput. Interact.}{5}{CSCW2}{}.
\newblock
\begin{APACrefURL} \url{https://doi.org/10.1145/3476046} \end{APACrefURL}
\newblock
\begin{APACrefDOI} \doi{10.1145/3476046} \end{APACrefDOI}
\PrintBackRefs{\CurrentBib}

\bibitem [\protect \citeauthoryear {%
Kim%
\ \BBA {} Priluck%
}{%
Kim%
\ \BBA {} Priluck%
}{%
{\protect \APACyear {2025}}%
}]{%
kim2025consumer}
\APACinsertmetastar {%
kim2025consumer}%
\begin{APACrefauthors}%
Kim, S.%
\BCBT {}\ \BBA {} Priluck, R.%
\end{APACrefauthors}%
\unskip\
\newblock
\APACrefYearMonthDay{2025}{}{}.
\newblock
{\BBOQ}\APACrefatitle {Consumer Responses to Generative AI Chatbots Versus Search Engines for Product Evaluation} {Consumer responses to generative ai chatbots versus search engines for product evaluation}.{\BBCQ}
\newblock
\APACjournalVolNumPages{Journal of Theoretical and Applied Electronic Commerce Research}{20}{2}{93}.
\PrintBackRefs{\CurrentBib}

\bibitem [\protect \citeauthoryear {%
Kojima%
, Gu%
, Reid%
, Matsuo%
\BCBL {}\ \BBA {} Iwasawa%
}{%
Kojima%
\ \protect \BOthers {.}}{%
{\protect \APACyear {2022}}%
}]{%
kojima2022large}
\APACinsertmetastar {%
kojima2022large}%
\begin{APACrefauthors}%
Kojima, T.%
, Gu, S\BPBI S.%
, Reid, M.%
, Matsuo, Y.%
\BCBL {}\ \BBA {} Iwasawa, Y.%
\end{APACrefauthors}%
\unskip\
\newblock
\APACrefYearMonthDay{2022}{}{}.
\newblock
{\BBOQ}\APACrefatitle {Large language models are zero-shot reasoners} {Large language models are zero-shot reasoners}.{\BBCQ}
\newblock
\APACjournalVolNumPages{Advances in neural information processing systems}{35}{}{22199--22213}.
\PrintBackRefs{\CurrentBib}

\bibitem [\protect \citeauthoryear {%
Kuhlthau%
}{%
Kuhlthau%
}{%
{\protect \APACyear {1991}}%
}]{%
kuhlthau1991inside}
\APACinsertmetastar {%
kuhlthau1991inside}%
\begin{APACrefauthors}%
Kuhlthau, C\BPBI C.%
\end{APACrefauthors}%
\unskip\
\newblock
\APACrefYearMonthDay{1991}{}{}.
\newblock
{\BBOQ}\APACrefatitle {Inside the search process: Information seeking from the user's perspective} {Inside the search process: Information seeking from the user's perspective}.{\BBCQ}
\newblock
\APACjournalVolNumPages{Journal of the American society for information science}{42}{5}{361--371}.
\PrintBackRefs{\CurrentBib}

\bibitem [\protect \citeauthoryear {%
Latour%
}{%
Latour%
}{%
{\protect \APACyear {1993}}%
}]{%
latour1993pasteurization}
\APACinsertmetastar {%
latour1993pasteurization}%
\begin{APACrefauthors}%
Latour, B.%
\end{APACrefauthors}%
\unskip\
\newblock
\APACrefYear{1993}.
\newblock
\APACrefbtitle {The pasteurization of France} {The pasteurization of france}.
\newblock
\APACaddressPublisher{}{Harvard University Press}.
\PrintBackRefs{\CurrentBib}

\bibitem [\protect \citeauthoryear {%
Laupichler%
, Aster%
, Schirch%
\BCBL {}\ \BBA {} Raupach%
}{%
Laupichler%
\ \protect \BOthers {.}}{%
{\protect \APACyear {2022}}%
}]{%
laupichler2022evaluation}
\APACinsertmetastar {%
laupichler2022evaluation}%
\begin{APACrefauthors}%
Laupichler, M\BPBI C.%
, Aster, A.%
, Schirch, J.%
\BCBL {}\ \BBA {} Raupach, T.%
\end{APACrefauthors}%
\unskip\
\newblock
\APACrefYearMonthDay{2022}{}{}.
\newblock
{\BBOQ}\APACrefatitle {Artificial intelligence literacy in higher and adult education: A scoping literature review} {Artificial intelligence literacy in higher and adult education: A scoping literature review}.{\BBCQ}
\newblock
\APACjournalVolNumPages{Computers and Education: Artificial Intelligence}{3}{}{100101}.
\PrintBackRefs{\CurrentBib}

\bibitem [\protect \citeauthoryear {%
Long%
\ \BBA {} Magerko%
}{%
Long%
\ \BBA {} Magerko%
}{%
{\protect \APACyear {2020}}%
}]{%
long2020ai}
\APACinsertmetastar {%
long2020ai}%
\begin{APACrefauthors}%
Long, D.%
\BCBT {}\ \BBA {} Magerko, B.%
\end{APACrefauthors}%
\unskip\
\newblock
\APACrefYearMonthDay{2020}{}{}.
\newblock
{\BBOQ}\APACrefatitle {What is AI Literacy? Competencies and Design Considerations} {What is ai literacy? competencies and design considerations}.{\BBCQ}
\newblock
\BIn{} \APACrefbtitle {Proceedings of the 2020 CHI Conference on Human Factors in Computing Systems} {Proceedings of the 2020 chi conference on human factors in computing systems}\ (\BPG~1–16).
\newblock
\APACaddressPublisher{New York, NY, USA}{Association for Computing Machinery}.
\newblock
\begin{APACrefURL} \url{https://doi.org/10.1145/3313831.3376727} \end{APACrefURL}
\newblock
\begin{APACrefDOI} \doi{10.1145/3313831.3376727} \end{APACrefDOI}
\PrintBackRefs{\CurrentBib}

\bibitem [\protect \citeauthoryear {%
Lupton%
}{%
Lupton%
}{%
{\protect \APACyear {2018}}%
}]{%
lupton2018data}
\APACinsertmetastar {%
lupton2018data}%
\begin{APACrefauthors}%
Lupton, D.%
\end{APACrefauthors}%
\unskip\
\newblock
\APACrefYearMonthDay{2018}{}{}.
\newblock
{\BBOQ}\APACrefatitle {How do data come to matter? Living and becoming with personal data} {How do data come to matter? living and becoming with personal data}.{\BBCQ}
\newblock
\APACjournalVolNumPages{Big Data \& Society}{5}{2}{2053951718786314}.
\PrintBackRefs{\CurrentBib}

\bibitem [\protect \citeauthoryear {%
Ma%
, Ng%
, Liu%
\BCBL {}\ \BBA {} Wong%
}{%
Ma%
\ \protect \BOthers {.}}{%
{\protect \APACyear {2025}}%
}]{%
ma2025fostering}
\APACinsertmetastar {%
ma2025fostering}%
\begin{APACrefauthors}%
Ma, M.%
, Ng, D\BPBI T\BPBI K.%
, Liu, Z.%
\BCBL {}\ \BBA {} Wong, G\BPBI K.%
\end{APACrefauthors}%
\unskip\
\newblock
\APACrefYearMonthDay{2025}{}{}.
\newblock
{\BBOQ}\APACrefatitle {Fostering Responsible AI Literacy: A Systematic Review of K-12 AI Ethics Education} {Fostering responsible ai literacy: A systematic review of k-12 ai ethics education}.{\BBCQ}
\newblock
\APACjournalVolNumPages{Computers and Education: Artificial Intelligence}{}{}{100422}.
\PrintBackRefs{\CurrentBib}

\bibitem [\protect \citeauthoryear {%
Maeda%
\ \BBA {} Quan-Haase%
}{%
Maeda%
\ \BBA {} Quan-Haase%
}{%
{\protect \APACyear {2024}}%
}]{%
maeda2024human}
\APACinsertmetastar {%
maeda2024human}%
\begin{APACrefauthors}%
Maeda, T.%
\BCBT {}\ \BBA {} Quan-Haase, A.%
\end{APACrefauthors}%
\unskip\
\newblock
\APACrefYearMonthDay{2024}{}{}.
\newblock
{\BBOQ}\APACrefatitle {When Human-AI Interactions Become Parasocial: Agency and Anthropomorphism in Affective Design} {When human-ai interactions become parasocial: Agency and anthropomorphism in affective design}.{\BBCQ}
\newblock
\BIn{} \APACrefbtitle {Proceedings of the 2024 ACM Conference on Fairness, Accountability, and Transparency} {Proceedings of the 2024 acm conference on fairness, accountability, and transparency}\ (\BPG~1068–1077).
\newblock
\APACaddressPublisher{New York, NY, USA}{Association for Computing Machinery}.
\newblock
\begin{APACrefURL} \url{https://doi.org/10.1145/3630106.3658956} \end{APACrefURL}
\newblock
\begin{APACrefDOI} \doi{10.1145/3630106.3658956} \end{APACrefDOI}
\PrintBackRefs{\CurrentBib}

\bibitem [\protect \citeauthoryear {%
Marchionini%
}{%
Marchionini%
}{%
{\protect \APACyear {2006}}%
}]{%
marchionini2006exploratory}
\APACinsertmetastar {%
marchionini2006exploratory}%
\begin{APACrefauthors}%
Marchionini, G.%
\end{APACrefauthors}%
\unskip\
\newblock
\APACrefYearMonthDay{2006}{}{}.
\newblock
{\BBOQ}\APACrefatitle {Exploratory search: from finding to understanding} {Exploratory search: from finding to understanding}.{\BBCQ}
\newblock
\APACjournalVolNumPages{Communications of the ACM}{49}{4}{41--46}.
\PrintBackRefs{\CurrentBib}

\bibitem [\protect \citeauthoryear {%
McDonald%
, Schoenebeck%
\BCBL {}\ \BBA {} Forte%
}{%
McDonald%
\ \protect \BOthers {.}}{%
{\protect \APACyear {2019}}%
}]{%
mcdonald_et_al_2019}
\APACinsertmetastar {%
mcdonald_et_al_2019}%
\begin{APACrefauthors}%
McDonald, N.%
, Schoenebeck, S.%
\BCBL {}\ \BBA {} Forte, A.%
\end{APACrefauthors}%
\unskip\
\newblock
\APACrefYearMonthDay{2019}{{\APACmonth{11}}}{}.
\newblock
{\BBOQ}\APACrefatitle {Reliability and Inter-rater Reliability in Qualitative Research: Norms and Guidelines for CSCW and HCI Practice} {Reliability and inter-rater reliability in qualitative research: Norms and guidelines for cscw and hci practice}.{\BBCQ}
\newblock
\APACjournalVolNumPages{Proc. ACM Hum.-Comput. Interact.}{3}{CSCW}{}.
\newblock
\begin{APACrefURL} \url{https://doi.org/10.1145/3359174} \end{APACrefURL}
\newblock
\begin{APACrefDOI} \doi{10.1145/3359174} \end{APACrefDOI}
\PrintBackRefs{\CurrentBib}

\bibitem [\protect \citeauthoryear {%
McHugh%
}{%
McHugh%
}{%
{\protect \APACyear {2012}}%
}]{%
mchugh2012interrater}
\APACinsertmetastar {%
mchugh2012interrater}%
\begin{APACrefauthors}%
McHugh, M\BPBI L.%
\end{APACrefauthors}%
\unskip\
\newblock
\APACrefYearMonthDay{2012}{}{}.
\newblock
{\BBOQ}\APACrefatitle {Interrater reliability: the kappa statistic} {Interrater reliability: the kappa statistic}.{\BBCQ}
\newblock
\APACjournalVolNumPages{Biochemia medica}{22}{3}{276--282}.
\PrintBackRefs{\CurrentBib}

\bibitem [\protect \citeauthoryear {%
Ng%
, Leung%
, Chu%
\BCBL {}\ \BBA {} Qiao%
}{%
Ng%
\ \protect \BOthers {.}}{%
{\protect \APACyear {2021}}%
}]{%
ng2021conceptualizing}
\APACinsertmetastar {%
ng2021conceptualizing}%
\begin{APACrefauthors}%
Ng, D\BPBI T\BPBI K.%
, Leung, J\BPBI K\BPBI L.%
, Chu, S\BPBI K\BPBI W.%
\BCBL {}\ \BBA {} Qiao, M\BPBI S.%
\end{APACrefauthors}%
\unskip\
\newblock
\APACrefYearMonthDay{2021}{}{}.
\newblock
{\BBOQ}\APACrefatitle {Conceptualizing AI literacy: An exploratory review} {Conceptualizing ai literacy: An exploratory review}.{\BBCQ}
\newblock
\APACjournalVolNumPages{Computers and Education: Artificial Intelligence}{2}{}{100041}.
\PrintBackRefs{\CurrentBib}

\bibitem [\protect \citeauthoryear {%
Ng%
, Wu%
, Leung%
, Chiu%
\BCBL {}\ \BBA {} Chu%
}{%
Ng%
\ \protect \BOthers {.}}{%
{\protect \APACyear {2024}}%
}]{%
ng2024design}
\APACinsertmetastar {%
ng2024design}%
\begin{APACrefauthors}%
Ng, D\BPBI T\BPBI K.%
, Wu, W.%
, Leung, J\BPBI K\BPBI L.%
, Chiu, T\BPBI K\BPBI F.%
\BCBL {}\ \BBA {} Chu, S\BPBI K\BPBI W.%
\end{APACrefauthors}%
\unskip\
\newblock
\APACrefYearMonthDay{2024}{}{}.
\newblock
{\BBOQ}\APACrefatitle {Design and validation of the AI literacy questionnaire: The affective, behavioural, cognitive and ethical approach} {Design and validation of the ai literacy questionnaire: The affective, behavioural, cognitive and ethical approach}.{\BBCQ}
\newblock
\APACjournalVolNumPages{British Journal of Educational Technology}{55}{3}{1082--1104}.
\PrintBackRefs{\CurrentBib}

\bibitem [\protect \citeauthoryear {%
Nissenbaum%
}{%
Nissenbaum%
}{%
{\protect \APACyear {2009}}%
}]{%
nissenbaum2010privacy}
\APACinsertmetastar {%
nissenbaum2010privacy}%
\begin{APACrefauthors}%
Nissenbaum, H.%
\end{APACrefauthors}%
\unskip\
\newblock
\APACrefYearMonthDay{2009}{}{}.
\newblock
{\BBOQ}\APACrefatitle {Privacy in context: Technology, policy, and the integrity of social life} {Privacy in context: Technology, policy, and the integrity of social life}.{\BBCQ}
\newblock
\BIn{} \APACrefbtitle {Privacy in context.} {Privacy in context.}
\newblock
\APACaddressPublisher{}{Stanford University Press}.
\PrintBackRefs{\CurrentBib}

\bibitem [\protect \citeauthoryear {%
O’Connor%
\ \BBA {} Joffe%
}{%
O’Connor%
\ \BBA {} Joffe%
}{%
{\protect \APACyear {2020}}%
}]{%
oconnor2020intercoder}
\APACinsertmetastar {%
oconnor2020intercoder}%
\begin{APACrefauthors}%
O’Connor, C.%
\BCBT {}\ \BBA {} Joffe, H.%
\end{APACrefauthors}%
\unskip\
\newblock
\APACrefYearMonthDay{2020}{}{}.
\newblock
{\BBOQ}\APACrefatitle {Intercoder reliability in qualitative research: Debates and practical guidelines} {Intercoder reliability in qualitative research: Debates and practical guidelines}.{\BBCQ}
\newblock
\APACjournalVolNumPages{International journal of qualitative methods}{19}{}{1609406919899220}.
\PrintBackRefs{\CurrentBib}

\bibitem [\protect \citeauthoryear {%
Pinch%
\ \BBA {} Bijker%
}{%
Pinch%
\ \BBA {} Bijker%
}{%
{\protect \APACyear {1984}}%
}]{%
pinch1984social}
\APACinsertmetastar {%
pinch1984social}%
\begin{APACrefauthors}%
Pinch, T\BPBI J.%
\BCBT {}\ \BBA {} Bijker, W\BPBI E.%
\end{APACrefauthors}%
\unskip\
\newblock
\APACrefYearMonthDay{1984}{}{}.
\newblock
{\BBOQ}\APACrefatitle {The social construction of facts and artefacts: Or how the sociology of science and the sociology of technology might benefit each other} {The social construction of facts and artefacts: Or how the sociology of science and the sociology of technology might benefit each other}.{\BBCQ}
\newblock
\APACjournalVolNumPages{Social studies of science}{14}{3}{399--441}.
\PrintBackRefs{\CurrentBib}

\bibitem [\protect \citeauthoryear {%
Pink%
, Sumartojo%
, Lupton%
\BCBL {}\ \BBA {} Heyes La~Bond%
}{%
Pink%
\ \protect \BOthers {.}}{%
{\protect \APACyear {2017}}%
}]{%
pink2017mundane}
\APACinsertmetastar {%
pink2017mundane}%
\begin{APACrefauthors}%
Pink, S.%
, Sumartojo, S.%
, Lupton, D.%
\BCBL {}\ \BBA {} Heyes La~Bond, C.%
\end{APACrefauthors}%
\unskip\
\newblock
\APACrefYearMonthDay{2017}{}{}.
\newblock
{\BBOQ}\APACrefatitle {Mundane data: The routines, contingencies and accomplishments of digital living} {Mundane data: The routines, contingencies and accomplishments of digital living}.{\BBCQ}
\newblock
\APACjournalVolNumPages{Big Data \& Society}{4}{1}{2053951717700924}.
\PrintBackRefs{\CurrentBib}

\bibitem [\protect \citeauthoryear {%
P{\"u}tz%
\ \BBA {} Esposito%
}{%
P{\"u}tz%
\ \BBA {} Esposito%
}{%
{\protect \APACyear {2024}}%
}]{%
putz2024performance}
\APACinsertmetastar {%
putz2024performance}%
\begin{APACrefauthors}%
P{\"u}tz, O.%
\BCBT {}\ \BBA {} Esposito, E.%
\end{APACrefauthors}%
\unskip\
\newblock
\APACrefYearMonthDay{2024}{}{}.
\newblock
{\BBOQ}\APACrefatitle {Performance without understanding: How ChatGPT relies on humans to repair conversational trouble} {Performance without understanding: How chatgpt relies on humans to repair conversational trouble}.{\BBCQ}
\newblock
\APACjournalVolNumPages{Discourse \& Communication}{18}{6}{859--868}.
\PrintBackRefs{\CurrentBib}

\bibitem [\protect \citeauthoryear {%
Ridley%
}{%
Ridley%
}{%
{\protect \APACyear {2024}}%
}]{%
ridley2024informing}
\APACinsertmetastar {%
ridley2024informing}%
\begin{APACrefauthors}%
Ridley, M.%
\end{APACrefauthors}%
\unskip\
\newblock
\APACrefYearMonthDay{2024}{}{}.
\newblock
{\BBOQ}\APACrefatitle {Informing Algorithmic Literacy Through User Folk Theories} {Informing algorithmic literacy through user folk theories}.{\BBCQ}
\newblock
\APACjournalVolNumPages{College \& Research Libraries}{85}{7}{966}.
\PrintBackRefs{\CurrentBib}

\bibitem [\protect \citeauthoryear {%
Ridley%
}{%
Ridley%
}{%
{\protect \APACyear {2025}}%
}]{%
ridley2025human}
\APACinsertmetastar {%
ridley2025human}%
\begin{APACrefauthors}%
Ridley, M.%
\end{APACrefauthors}%
\unskip\
\newblock
\APACrefYearMonthDay{2025}{}{}.
\newblock
{\BBOQ}\APACrefatitle {Human-centered explainable artificial intelligence: An Annual Review of Information Science and Technology (ARIST) paper} {Human-centered explainable artificial intelligence: An annual review of information science and technology (arist) paper}.{\BBCQ}
\newblock
\APACjournalVolNumPages{Journal of the Association for Information Science and Technology}{76}{1}{98--120}.
\PrintBackRefs{\CurrentBib}

\bibitem [\protect \citeauthoryear {%
Rieh%
}{%
Rieh%
}{%
{\protect \APACyear {2002}}%
}]{%
rieh2002judgment}
\APACinsertmetastar {%
rieh2002judgment}%
\begin{APACrefauthors}%
Rieh, S\BPBI Y.%
\end{APACrefauthors}%
\unskip\
\newblock
\APACrefYearMonthDay{2002}{}{}.
\newblock
{\BBOQ}\APACrefatitle {Judgment of information quality and cognitive authority in the Web} {Judgment of information quality and cognitive authority in the web}.{\BBCQ}
\newblock
\APACjournalVolNumPages{Journal of the American society for information science and technology}{53}{2}{145--161}.
\PrintBackRefs{\CurrentBib}

\bibitem [\protect \citeauthoryear {%
Salovaara%
, Helfenstein%
\BCBL {}\ \BBA {} Oulasvirta%
}{%
Salovaara%
\ \protect \BOthers {.}}{%
{\protect \APACyear {2011}}%
}]{%
salovaara2011everyday}
\APACinsertmetastar {%
salovaara2011everyday}%
\begin{APACrefauthors}%
Salovaara, A.%
, Helfenstein, S.%
\BCBL {}\ \BBA {} Oulasvirta, A.%
\end{APACrefauthors}%
\unskip\
\newblock
\APACrefYearMonthDay{2011}{}{}.
\newblock
{\BBOQ}\APACrefatitle {Everyday appropriations of information technology: A study of creative uses of digital cameras} {Everyday appropriations of information technology: A study of creative uses of digital cameras}.{\BBCQ}
\newblock
\APACjournalVolNumPages{Journal of the American Society for Information Science and Technology}{62}{12}{2347--2363}.
\PrintBackRefs{\CurrentBib}

\bibitem [\protect \citeauthoryear {%
Saracevic%
}{%
Saracevic%
}{%
{\protect \APACyear {2007}}%
}]{%
saracevic2007relevance}
\APACinsertmetastar {%
saracevic2007relevance}%
\begin{APACrefauthors}%
Saracevic, T.%
\end{APACrefauthors}%
\unskip\
\newblock
\APACrefYearMonthDay{2007}{}{}.
\newblock
{\BBOQ}\APACrefatitle {Relevance: A review of the literature and a framework for thinking on the notion in information science. Part II: Nature and manifestations of relevance} {Relevance: A review of the literature and a framework for thinking on the notion in information science. part ii: Nature and manifestations of relevance}.{\BBCQ}
\newblock
\APACjournalVolNumPages{Journal of the American society for information science and technology}{58}{13}{1915--1933}.
\PrintBackRefs{\CurrentBib}

\bibitem [\protect \citeauthoryear {%
Schulz-Schaeffer%
}{%
Schulz-Schaeffer%
}{%
{\protect \APACyear {2025}}%
}]{%
schulz2025generative}
\APACinsertmetastar {%
schulz2025generative}%
\begin{APACrefauthors}%
Schulz-Schaeffer, I.%
\end{APACrefauthors}%
\unskip\
\newblock
\APACrefYearMonthDay{2025}{}{}.
\newblock
{\BBOQ}\APACrefatitle {Why generative AI is different from designed technology regarding task-relatedness, user interaction, and agency} {Why generative ai is different from designed technology regarding task-relatedness, user interaction, and agency}.{\BBCQ}
\newblock
\APACjournalVolNumPages{Big Data \& Society}{12}{3}{20539517251367452}.
\PrintBackRefs{\CurrentBib}

\bibitem [\protect \citeauthoryear {%
Sebastian%
}{%
Sebastian%
}{%
{\protect \APACyear {2025}}%
}]{%
sebastian2025reframing}
\APACinsertmetastar {%
sebastian2025reframing}%
\begin{APACrefauthors}%
Sebastian, J\BPBI K.%
\end{APACrefauthors}%
\unskip\
\newblock
\APACrefYearMonthDay{2025}{}{}.
\newblock
{\BBOQ}\APACrefatitle {Reframing information seeking in the age of generative AI: A critical and humanistic approach} {Reframing information seeking in the age of generative ai: A critical and humanistic approach}.{\BBCQ}
\newblock
\BIn{} \APACrefbtitle {ACRL 2025 conference} {Acrl 2025 conference}\ (\BPG~527).
\PrintBackRefs{\CurrentBib}

\bibitem [\protect \citeauthoryear {%
Shah%
\ \BBA {} Bender%
}{%
Shah%
\ \BBA {} Bender%
}{%
{\protect \APACyear {2022}}%
}]{%
shah_bender_22}
\APACinsertmetastar {%
shah_bender_22}%
\begin{APACrefauthors}%
Shah, C.%
\BCBT {}\ \BBA {} Bender, E\BPBI M.%
\end{APACrefauthors}%
\unskip\
\newblock
\APACrefYearMonthDay{2022}{}{}.
\newblock
{\BBOQ}\APACrefatitle {Situating Search} {Situating search}.{\BBCQ}
\newblock
\BIn{} \APACrefbtitle {Proceedings of the 2022 Conference on Human Information Interaction and Retrieval} {Proceedings of the 2022 conference on human information interaction and retrieval}\ (\BPG~221–232).
\newblock
\APACaddressPublisher{New York, NY, USA}{Association for Computing Machinery}.
\newblock
\begin{APACrefURL} \url{https://doi.org/10.1145/3498366.3505816} \end{APACrefURL}
\newblock
\begin{APACrefDOI} \doi{10.1145/3498366.3505816} \end{APACrefDOI}
\PrintBackRefs{\CurrentBib}

\bibitem [\protect \citeauthoryear {%
Silverstone%
}{%
Silverstone%
}{%
{\protect \APACyear {2006}}%
}]{%
silverstone2006domesticating}
\APACinsertmetastar {%
silverstone2006domesticating}%
\begin{APACrefauthors}%
Silverstone, R.%
\end{APACrefauthors}%
\unskip\
\newblock
\APACrefYearMonthDay{2006}{}{}.
\newblock
{\BBOQ}\APACrefatitle {Domesticating domestication: Reflections on the life of a concept} {Domesticating domestication: Reflections on the life of a concept}.{\BBCQ}
\newblock
\BIn{} T.~Berker, M.~Hartmann, Y.~Punie\BCBL {}\ \BBA {} K.~Ward\ (\BEDS), \APACrefbtitle {Domestication of Media and Technology} {Domestication of media and technology}\ (\BPGS\ 229--248).
\newblock
\APACaddressPublisher{}{Open University Press}.
\PrintBackRefs{\CurrentBib}

\bibitem [\protect \citeauthoryear {%
Silverstone%
, Hirsch%
\BCBL {}\ \BBA {} Morley%
}{%
Silverstone%
\ \protect \BOthers {.}}{%
{\protect \APACyear {1994}}%
}]{%
silverstone1992information}
\APACinsertmetastar {%
silverstone1992information}%
\begin{APACrefauthors}%
Silverstone, R.%
, Hirsch, E.%
\BCBL {}\ \BBA {} Morley, D.%
\end{APACrefauthors}%
\unskip\
\newblock
\APACrefYearMonthDay{1994}{}{}.
\newblock
{\BBOQ}\APACrefatitle {Information and communication technologies and the moral economy of the household} {Information and communication technologies and the moral economy of the household}.{\BBCQ}
\newblock
\BIn{} R.~Silverstone\ \BBA {} E.~Hirsch\ (\BEDS), \APACrefbtitle {Consuming Technologies: Media and Information in Domestic Spaces} {Consuming technologies: Media and information in domestic spaces}\ (\BPGS\ 13--28).
\newblock
\APACaddressPublisher{}{Routledge}.
\PrintBackRefs{\CurrentBib}

\bibitem [\protect \citeauthoryear {%
S{\o}rensen%
}{%
S{\o}rensen%
}{%
{\protect \APACyear {2006}}%
}]{%
sorensen2006domestication}
\APACinsertmetastar {%
sorensen2006domestication}%
\begin{APACrefauthors}%
S{\o}rensen, K\BPBI H.%
\end{APACrefauthors}%
\unskip\
\newblock
\APACrefYearMonthDay{2006}{}{}.
\newblock
{\BBOQ}\APACrefatitle {Domestication: the enactment of technology} {Domestication: the enactment of technology}.{\BBCQ}
\newblock
\APACjournalVolNumPages{Domestication of media and technology}{46}{}{}.
\PrintBackRefs{\CurrentBib}

\bibitem [\protect \citeauthoryear {%
Tomisu%
, Ueda%
\BCBL {}\ \BBA {} Yamanaka%
}{%
Tomisu%
\ \protect \BOthers {.}}{%
{\protect \APACyear {2025}}%
}]{%
tomisu2025cognitive}
\APACinsertmetastar {%
tomisu2025cognitive}%
\begin{APACrefauthors}%
Tomisu, H.%
, Ueda, J.%
\BCBL {}\ \BBA {} Yamanaka, T.%
\end{APACrefauthors}%
\unskip\
\newblock
\APACrefYearMonthDay{2025}{}{}.
\newblock
{\BBOQ}\APACrefatitle {The cognitive mirror: A framework for AI-powered metacognition and self-regulated learning} {The cognitive mirror: A framework for ai-powered metacognition and self-regulated learning}.{\BBCQ}
\newblock
\BIn{} \APACrefbtitle {Frontiers in Education} {Frontiers in education}\ (\BVOL~10, \BPG~1697554).
\PrintBackRefs{\CurrentBib}

\bibitem [\protect \citeauthoryear {%
Touretzky%
, Gardner-McCune%
\BCBL {}\ \BBA {} Seehorn%
}{%
Touretzky%
\ \protect \BOthers {.}}{%
{\protect \APACyear {2023}}%
}]{%
touretzky2023machine}
\APACinsertmetastar {%
touretzky2023machine}%
\begin{APACrefauthors}%
Touretzky, D.%
, Gardner-McCune, C.%
\BCBL {}\ \BBA {} Seehorn, D.%
\end{APACrefauthors}%
\unskip\
\newblock
\APACrefYearMonthDay{2023}{}{}.
\newblock
{\BBOQ}\APACrefatitle {Machine learning and the five big ideas in AI} {Machine learning and the five big ideas in ai}.{\BBCQ}
\newblock
\APACjournalVolNumPages{International journal of artificial intelligence in education}{33}{2}{233--266}.
\PrintBackRefs{\CurrentBib}

\bibitem [\protect \citeauthoryear {%
Urquhart%
, Cheuk%
, Lam%
\BCBL {}\ \BBA {} Snowden%
}{%
Urquhart%
\ \protect \BOthers {.}}{%
{\protect \APACyear {2025}}%
}]{%
urquhart2025sense}
\APACinsertmetastar {%
urquhart2025sense}%
\begin{APACrefauthors}%
Urquhart, C.%
, Cheuk, B.%
, Lam, L.%
\BCBL {}\ \BBA {} Snowden, D.%
\end{APACrefauthors}%
\unskip\
\newblock
\APACrefYearMonthDay{2025}{}{}.
\newblock
{\BBOQ}\APACrefatitle {Sense-making, sensemaking and sense making—A systematic review and meta-synthesis of literature in information science and education: An Annual Review of Information Science and Technology (ARIST) paper} {Sense-making, sensemaking and sense making—a systematic review and meta-synthesis of literature in information science and education: An annual review of information science and technology (arist) paper}.{\BBCQ}
\newblock
\APACjournalVolNumPages{Journal of the Association for Information Science and Technology}{76}{1}{3--97}.
\PrintBackRefs{\CurrentBib}

\bibitem [\protect \citeauthoryear {%
Wei%
\ \protect \BOthers {.}}{%
Wei%
\ \protect \BOthers {.}}{%
{\protect \APACyear {2022}}%
}]{%
wei2022chain}
\APACinsertmetastar {%
wei2022chain}%
\begin{APACrefauthors}%
Wei, J.%
, Wang, X.%
, Schuurmans, D.%
, Bosma, M.%
, Xia, F.%
, Chi, E.%
\BDBL {}others%
\end{APACrefauthors}%
\unskip\
\newblock
\APACrefYearMonthDay{2022}{}{}.
\newblock
{\BBOQ}\APACrefatitle {Chain-of-thought prompting elicits reasoning in large language models} {Chain-of-thought prompting elicits reasoning in large language models}.{\BBCQ}
\newblock
\APACjournalVolNumPages{Advances in neural information processing systems}{35}{}{24824--24837}.
\PrintBackRefs{\CurrentBib}

\bibitem [\protect \citeauthoryear {%
Wilson%
}{%
Wilson%
}{%
{\protect \APACyear {1999}}%
}]{%
wilson1999models}
\APACinsertmetastar {%
wilson1999models}%
\begin{APACrefauthors}%
Wilson, T\BPBI D.%
\end{APACrefauthors}%
\unskip\
\newblock
\APACrefYearMonthDay{1999}{}{}.
\newblock
{\BBOQ}\APACrefatitle {Models in information behaviour research} {Models in information behaviour research}.{\BBCQ}
\newblock
\APACjournalVolNumPages{Journal of documentation}{55}{3}{249--270}.
\PrintBackRefs{\CurrentBib}

\bibitem [\protect \citeauthoryear {%
Zhang%
\ \protect \BOthers {.}}{%
Zhang%
\ \protect \BOthers {.}}{%
{\protect \APACyear {2024}}%
}]{%
zhang2024fairgame}
\APACinsertmetastar {%
zhang2024fairgame}%
\begin{APACrefauthors}%
Zhang, Z.%
, Jia, M.%
, Lee, H\BHBI P\BPBI H.%
, Yao, B.%
, Das, S.%
, Lerner, A.%
\BDBL {}Li, T.%
\end{APACrefauthors}%
\unskip\
\newblock
\APACrefYearMonthDay{2024}{}{}.
\newblock
{\BBOQ}\APACrefatitle {“It's a Fair Game”, or Is It? Examining How Users Navigate Disclosure Risks and Benefits When Using LLM-Based Conversational Agents} {“it's a fair game”, or is it? examining how users navigate disclosure risks and benefits when using llm-based conversational agents}.{\BBCQ}
\newblock
\BIn{} \APACrefbtitle {Proceedings of the 2024 CHI Conference on Human Factors in Computing Systems.} {Proceedings of the 2024 chi conference on human factors in computing systems.}
\newblock
\APACaddressPublisher{New York, NY, USA}{Association for Computing Machinery}.
\newblock
\begin{APACrefURL} \url{https://doi.org/10.1145/3613904.3642385} \end{APACrefURL}
\newblock
\begin{APACrefDOI} \doi{10.1145/3613904.3642385} \end{APACrefDOI}
\PrintBackRefs{\CurrentBib}

\end{thebibliography}
\end{document}